%% file: main.tex
\documentclass{tlp}

\usepackage{epsfig}

\usepackage{graphicx}
\usepackage[all]{xy}
\usepackage{eepic,latexsym,amssymb}
\usepackage{fancybox,epic}
\usepackage{algorithm}
\usepackage[noend]{algpseudocode}

\newtheorem{theorem}{Theorem}[section]
\newtheorem{proposition}[theorem]{Proposition}
\newtheorem{example}[theorem]{Example}
\newtheorem{definition}[theorem]{Definition}

\newtheorem{corollary}[theorem]{Corollary}

\newcommand{\vars}{{\cal V}}
\newcommand{\entries}{{\it Entry}}
\newcommand{\depend}{{\it Dep}}

\newcommand{\prog}{{\it Prog}}

\newcommand{\alub}{\mbox{\sf Alub}}

\newcommand{\strategy}{\Omega}

\newcommand{\adom}{{\it ADom}}
\newcommand{\aint}{{\it APol}}
\newcommand{\aatom}{{\it AA}\-{\it tom}}
\newcommand{\acert}{{\it ACert}}
\newcommand{\dom}{D_\alpha}
\newcommand{\inten}{I_\alpha}
\newcommand{\atom}{S_\alpha}
\newcommand{\cert}{Cert_\alpha}

\newcommand{\qhs}{{\it QHS}}

\newcommand{\lsem}{\mbox{$\lbrack\hspace{-0.3ex}\lbrack$}}
\newcommand{\rsem}{\mbox{$\rbrack\hspace{-0.3ex}\rbrack$}}
\newcommand{\sqbrack}[1]{\lsem #1 \rsem}

\newcommand{\p}{{\sqbrack{P}}}

\newcommand{\semabs}{\p_\alpha}

\newcommand{\CP}{\mbox{\it CP}}

\newcommand{\AP}{\mbox{\it AP}}

\newcommand{\short}[1]{}

\long\def\comment#1{}

\def\tuple#1{\langle{#1}\rangle}

\newcommand{\ciao}{\texttt{Ciao}}
\newcommand{\ciaopp}{\texttt{CiaoPP}}

\newcommand{\analyzerA}{{\sc An}\-{\sc a}\-{\sc ly}\-{\sc ze\_f}}
\newcommand{\certifierA}{{\sc Cer}\-{\sc ti}\-{\sc fier\_f}}
\newcommand{\checkerA}{{\sc Checker\_f}}

\newcommand{\certificateA}{{\sf FCert}}

\newcommand{\analyzerB}{{\sc An}\-{\sc a}\-{\sc ly}\-{\sc ze\_r}}
\newcommand{\certifierB}{{\sc Certifier\_r}}
\newcommand{\checkerB}{{\sc Checker\_r}}
\newcommand{\certificateB}{{\sf RCert}}
\newcommand{\checkingB}{{\sc Checking\_r}}

\newcommand{\error}{{\sf error}}
\newcommand{\true}{{\sf true}}

\newcommand{\false}{{\sf false}}

\newcommand{\reduced}{\sf RED}

\renewcommand{\intextsep}{2pt}
\renewcommand{\floatsep}{4pt}

\title[Theory and Practice of Logic Programming]
        {Certificate Size Reduction in Abstraction-Carrying
          Code\thanks{A preliminary
    version  of this work appeared in the Proceedings of ICLP'06
    (Albert et al. 2006).}}
\author[E. Albert, P. Arenas, G. Puebla and M. Hermenegildo]
         {ELVIRA ALBERT, PURI ARENAS\\
School of Computer Science, Complutense University of Madrid \\
 E28040-Profesor Jos\'e Garc\'{\i}a Santesmases, s/n, Madrid,
 Spain \\
         \email{\{elvira,puri\}@sip.ucm.es}
         \and GERM\'AN PUEBLA$^1$, MANUEL HERMENEGILDO$^{1,2}$ \\
$^1$ School of Computer Science, Technical University of Madrid\\
E28660-Boadilla del Monte, Madrid, Spain \\
         \email{\{german,herme\}@fi.upm.es} \\
$^2$ Madrid Institute for Advanced Studies\\ 
in Software Development Technology (IMDEA Software)\\
    Madrid,Spain\\
    \email{manuel.hermenegildo@imdea.org}}

\begin{document}

\label{firstpage}

\maketitle

\begin{abstract}
  \emph{Abstraction-Carrying Code} (ACC) has recently been proposed as
  a framework for mobile code safety in which the code supplier
  provides a program together with an \emph{abstraction} (or abstract
  model of the program) whose validity entails compliance with a
  predefined safety policy. The abstraction plays thus the role of
  safety certificate and its generation is carried out automatically
  by a fixpoint analyzer. The advantage of providing a (fixpoint)
  abstraction to the code consumer is that its validity is checked in
  a \emph{single pass} (i.e., one iteration) of an abstract
  interpretation-based checker.  A main challenge to make ACC useful
  in practice is to reduce the size of certificates as much as
  possible while at the same time not increasing checking time. The
  intuitive idea is to only include in the certificate information
  that the checker is unable to reproduce without iterating. We
  introduce the notion of \emph{reduced certificate} which
  characterizes the subset of the abstraction which a checker needs in
  order to validate (and re-construct) the \emph{full certificate} in
  a single pass.  Based on this notion, we instrument a generic
  analysis algorithm with the necessary extensions in order to
  identify the information relevant to the checker.  Interestingly,
  the fact that the reduced certificate omits (parts of) the
  abstraction has implications in the design of the checker. We
  provide the sufficient conditions which allow us to ensure that 1)
  if the checker succeeds in validating the certificate, then the
  certificate is valid for the program (correctness) and 2) the
  checker will succeed for any reduced certificate which is valid
  (completeness).  Our approach has been implemented and benchmarked
  within the \ciaopp~system. The experimental results show that our
  proposal is able to greatly reduce the size of certificates in
  practice.

\noindent
{\em To appear in Theory and Practice of Logic Programming (TPLP).}

\end{abstract}

\begin{keywords}
Proof-Carrying Code. Abstraction-Carrying Code. Static
Analysis. Reduced Certificates.
\end{keywords}

\pagestyle{plain}

\input intro

\input foundations

\input example

\input analysis

\input redundant

\input updates

\input checking

\input experiments

\input related

\section*{Acknowledgments}

The authors would like to gratefully thank the anonymous referees 
for useful comments on a preliminary version of this article.
This work was funded in part by the Information \& Communication
Technologies program of the European Commission, Future and Emerging
Technologies (FET), under the ICT-231620 {\em HATS} project, by the
Spanish Ministry of Science and Innovation (MICINN) under the
TIN-2008-05624 {\em DOVES} project, the 
TIN2008-04473-E (Acci\'on Especial) project, the HI2008-0153 (Acci\'on
Integrada) project, the UCM-BSCH-GR58/08-910502 Research Group
and by the Madrid Regional Government under the
S2009TIC-1465 \emph{PROMETIDOS} project.

\bibliographystyle{acmtrans}

\end{document}

%% file: intro.tex

\section{Introduction}\label{sec:introduction}

Proof-Carrying Code (PCC) \cite{Nec97} is a general framework for
mobile code safety which proposes to associate safety information in
the form of a \emph{certificate} to programs. The certificate (or
proof) is created at compile time by the \emph{certifier} on
the code supplier side, and it is packaged along with the 
code.  The consumer which receives or downloads the (untrusted)
code+certificate 
package can then run a \emph{checker} which by an efficient inspection
of the code and the certificate can verify the validity of the
certificate and thus compliance with the safety policy.  The key
benefit of this ``certificate-based'' approach to mobile code safety
is that the task of the consumer is reduced from the level of proving to
the level of checking, a procedure that should be much simpler, efficient,
and automatic than generating the original certificate.

Abstraction-Carrying Code
(ACC)~\cite{lpar04-ai-safety-long,ai-safety-ngc07} has been  
recently proposed as an enabling technology for PCC in which an
\emph{abstraction} (or abstract model of the program) plays the role
of certificate. An important feature of ACC is that not only the
checking, but also the generation of the abstraction, is carried out
automatically by a fixpoint analyzer.  In this article we will
consider analyzers which construct a program \emph{analysis graph}
which is interpreted as an abstraction of the (possibly infinite) set
of states explored by the concrete execution.  To capture the
different graph traversal strategies used in different fixpoint
algorithms, we use the \emph{generic} description of
\cite{incanal-toplas}, which generalizes the algorithms used in
state-of-the-art analysis engines.

Essentially, the
certification/analysis carried out by the supplier is an iterative
process which repeatedly traverses the analysis graph until a fixpoint
is reached.  The analysis information inferred for each call which
appears during the (multiple) graph traversals is stored in the
\emph{answer table} \cite{incanal-toplas}. After each iteration (or
graph traversal), if the answer computed for a certain call is
different from the one previously stored in the answer table, both
answers are combined (by computing their lub) and the result is used
1) to \emph{update} the table, and 2) to launch the recomputation of
those calls whose answer depends on the answer currently computed.  
In the original ACC
framework, the final \emph{full} answer table constitutes the
certificate. A main idea is that, since this certificate contains the
fixpoint, a single pass over the analysis graph is sufficient to
validate such certificate on the consumer side.

One of the main challenges for the practical uptake of ACC (and
related methods) is to produce certificates which are reasonably
small.  This is important 
since the certificate is transmitted together with the untrusted code
and, hence, reducing its size will presumably contribute to a smaller
transmission time --very relevant for instance under 
limited bandwidth and/or expensive network connectivity conditions.  
Also, this reduces the storage cost for the certificate.
Nevertheless, a main concern when reducing the size of the certificate
is that checking time is not increased
(among other reasons because pervasive and embedded systems also 
suffer typically from limited computing --and power-- resources).
In principle, the consumer could use an analyzer for the purpose of
generating the whole fixpoint from scratch, which is still feasible as
analysis is automatic. However, this would defeat one of the main
purposes of ACC, which is to reduce checking time. The objective of
this work is to characterize the smallest subset of the abstraction
which must be sent within a certificate --and which still guarantees a
single pass checking process-- and to design an ACC scheme which
generates and validates such reduced certificates.
The main contributions of this article are:

\begin{enumerate}
\item The notion of \emph{reduced certificate} which characterizes the
  subset of the abstraction which, for a given analysis graph
  traversal strategy, the checker needs in order to validate (and
  re-construct) the full certificate in a single pass.

\item An instrumentation of the generic abstract interpretation-based
analysis algorithm of \cite{incanal-toplas} with the necessary
extensions in order to identify relevant information to the checker.

\item A checker for reduced certificates which is \emph{correct},
  i.e., if the checker succeeds in validating the certificate, then
  the certificate is valid for the program.

\item Sufficient conditions for ensuring
  \emph{completeness} of the checking process. Concretely, if the
  checker uses the same strategy as the analyzer then our proposed
  checker will succeed for any reduced certificate which is
  valid.

\item An experimental evaluation of the effect of our approach on the
  \ciaopp\ system \cite{ciaopp-sas03-journal-scp}, the abstract
  interpretation-based preprocessor of the \ciao~multi-paradigm
  (Constraint) Logic Programming system. The experimental results show
  that the certificate can be greatly reduced (by a factor of 3.35)
  with no  increase in checking time.
 \end{enumerate}

Both the ACC framework and our work here are applied at 
the \emph{source} level. In contrast, in existing PCC
frameworks, the code supplier typically packages the certificate with
the \emph{object} code rather than with the \emph{source} code (both
are untrusted). Nevertheless, our choice of making our presentation 
at the source level is without loss of generality
because both the original ideas in the ACC approach and those in
our current proposal 
can also be applied directly to bytecode.  Indeed, a good number of abstract
interpretation-based analyses have been proposed in the literature for
bytecode and machine code, most of which compute a fixpoint during
analysis which can be reduced using the general principle of our
proposal.  For instance, in recent work, the concrete CLP verifier
used in the original ACC implementation has itself been shown to be
applicable without modification also to Java bytecode via a
transformational approach, based on partial
evaluation~\cite{jvm-pe-padl07}
or via direct transformation~\cite{decomp-oo-prolog-lopstr07} using
standard tools such as Soot~\cite{vall99soot}. Furthermore,
in~\cite{fixpt-javabytecode-bytecode07,decomp-oo-prolog-lopstr07} a
fixpoint-based analysis framework has been developed specifically for
Java bytecode which is essentially equivalent to that used in 
the ACC proposal and to the one that we will
apply in this work on the producer side to perform the analysis and
verification.  This supports the direct 
applicability of our approach to bytecode-based program
representations and, in general, to other languages and paradigms.

The rest of the article is organized as follows. The following section
presents a general view of ACC. Section \ref{sec:informal} gives 
a brief overview of our method by means of a simple example.
Section~\ref{sec:an-abstr-interpr} recalls the
certification process performed by the code supplier and illustrates
it with a running example. Section~\ref{sec:reduced-certificates} 
characterizes the notion of reduced certificate and instruments a
generic certifier for its generation.
Section~\ref{sec:generic-checker-full} presents a generic checker for
reduced certificates together with correctness and completeness
results. Finally, Section~\ref{sec:experiments} discusses some
experimental results and related work.

%% file: foundations.tex

\section{A General View of Abstraction-Carrying Code}\label{sec:basics-abstr-carry}

We assume the reader familiar with abstract interpretation~(see
\cite{Cousot77})  and (Constraint) Logic Programming (C)LP~(see, e.g.,
\cite{marriot-stuckey-98} and \cite{Lloyd87}).

An abstract interpretation-based certifier is a function ${\sf
  certifier}: \prog \times \adom \times \aint \mapsto \acert$ which
for a given program $P \in \prog$, an abstract domain $\tuple{\dom,
  \sqsubseteq} \in \adom$ and a safety policy $\inten \in \aint$
generates a certificate $\cert \in \acert$, by using an abstract
interpreter for $\dom$, which entails that $P$ satisfies $\inten$.
In the following, we denote that $\inten$ and $\cert$ are
specifications given as abstract semantic values of $\dom$ by using
the same $\alpha$.
The essential idea in the certification process carried out in ACC is
that a fixpoint static analyzer is used to
automatically 
infer an abstract model (or simply \emph{abstraction}) about the
mobile code which can then be used to prove that the code is safe
w.r.t.\ the given policy in a straightforward way. The basics for
defining the abstract interpretation-based certifiers in ACC are
summarized in the following four points and equations.

\begin{description}

\item[\em Approximation.] We consider a {\em description (or abstract) domain} $\langle \dom,
{\sqsubseteq} \rangle \in \adom$ and its corresponding \emph{concrete
  domain} $\langle 2^D, \subseteq \rangle$, both with a complete
lattice structure.  Description (or abstract) values and sets of
concrete values are related by an {\em abstraction} function $\alpha:
2^{D}\rightarrow \dom$, and a {\em concretization} function $\gamma:
\dom\rightarrow 2^{D}$.  The pair $\langle \alpha, \gamma \rangle$
forms a Galois connection.  The concrete and abstract domains must be
related in such a way that the following condition
holds~\cite{Cousot77}:

\[\forall x{\in} 2^D,\forall y{\in}
  \dom:~
  (\alpha(x) \sqsubseteq y) \Longleftrightarrow(x\subseteq \gamma(y))\]

\noindent
In general ${\sqsubseteq}$ is induced by $\subseteq$ and $\alpha$.
Similarly, the operations of {\em least upper bound\/} ($\sqcup$) and
{\em greatest lower bound\/} ($\sqcap$) mimic those of $2^D$ in a
precise sense.

\medskip

\item[\em Abstraction generation.] We consider the class of {\em
    fixpoint semantics} in which a (monotonic) semantic operator,
  $S_P$, is associated to each program $P$.
  The meaning of the program, $\lsem P \rsem$,
  is defined as the least fixed point of the $S_P$ operator, i.e.,
  $\lsem P \rsem {=} {\rm lfp}(S_P)$.  If $S_P$ is continuous, the least
  fixed point is the limit of an iterative process involving at most
  $\omega$ applications of $S_P$ starting from the bottom element
  of the lattice. Using abstract interpretation, we can 
  use an operator $S_P^\alpha$ which works in the abstract domain and
  which is the abstract counterpart of $S_P$. This operator induces
  the abstract meaning of the program, which we refer to as
  $\p_\alpha$. Now, again, starting from the bottom element of the
  lattice we can obtain the least fixpoint of $S_P^\alpha$, denoted
  ${\rm lfp}(S_P^\alpha)$, and we define $\p_\alpha{=}{\rm
    lfp}(S_P^\alpha)$. 
  Correctness of analysis~\cite{Cousot77} ensures that $\semabs$
  safely approximates $\sqbrack{P}$, i.e., $\sqbrack{P} \in
  \gamma(\sqbrack{P}_\alpha)$. In actual analyzers, it is often the
  case that the analysis computes a post-fixpoint of $S_P^\alpha$, which
  we refer to as $\cert$, instead of the least fixpoint. The reason
  for this is that computing the least fixpoint may require a too
  large (even infinite) number of iterations. 
  An analyzer is a function ${\sf analyzer}:{\it Prog} \times \adom
  \mapsto \acert$ such that:
\begin{equation}\label{eq:1}
{\sf analyzer}(P,\dom) {=} \cert \wedge S_P^\alpha(\cert) {=} \cert 
\end{equation}

\noindent
Since $\sqbrack{P}_\alpha\sqsubseteq\cert$, $\cert$ is a safe
approximation of $\lsem P \rsem$.

\medskip

\item[\em Verification Condition.]
Let $\cert$ be a safe approximation of $\lsem P \rsem$.
  If an abstract safety specification $\inten$ can be proved w.r.t.\
  $\cert$, then $P$ satisfies the safety
  policy and $\cert$ is a valid certificate:

\begin{equation}\label{eq:2}
\cert \mbox{ is \emph{a valid certificate} for $P$ w.r.t. }
 \inten \mbox{ if }
    \cert \sqsubseteq \inten
\end{equation}

\medskip

\item[\em Certification.] Together, Equations (\ref{eq:1}) and (\ref{eq:2})
  define a certifier which provides program fixpoints, $ \cert$, as certificates
  which entail a given safety policy, i.e., by taking $\cert  =  {\sf analyzer}(P,\dom)$.
\end{description}
The second main idea in ACC is that a simple, easy-to-trust abstract
interpretation-based checker verifies the validity of the abstraction
on the mobile code.  The checker is defined as a specialized abstract
interpreter whose key characteristic is that it does not need to
iterate in order to reach a fixpoint (in contrast to standard
analyzers).  The basics for defining the abstract interpretation-based
checkers in ACC are summarized in the following two points and
equations.
\begin{description}

\item[\em Checking.]
If a certificate $\cert$ is a fixpoint of $S^{\alpha}_P$, then
    $S^{\alpha}_P(\cert) = \cert$. Thus, a checker is a function ${\sf
    checker}:{\it Prog} \times \adom \times \acert \mapsto {\it bool}$
  which for a program $P {\in} {\it Prog}$, an abstract domain $\dom {\in}
    \adom$ and a
  certificate $\cert {\in} \acert$ checks
whether $\cert$ is a fixpoint of $S^{\alpha}_P$ or not:
\begin{equation}\label{eq:3}
{\sf checker}(P,\dom,\cert) \ {\it returns\  true\ iff}\  (S^{\alpha}_P(\cert) = \cert)
\end{equation}

\medskip

\item[\em Verification Condition Regeneration.]  
To retain the safety
  guarantees, the consumer must regenerate a trustworthy verification
  condition --Equation \ref{eq:2}-- and use the incoming certificate
  to test for adherence to the  safety policy.

\begin{equation}\label{eq:4}
{\it  P \ is \ trusted\ iff}\  \cert \sqsubseteq \inten
\end{equation}
\end{description}

\noindent
Therefore, the general idea in ACC is that, while analysis --Equation
(\ref{eq:1})-- is an iterative process, which may traverse (parts of)
the abstraction more than once until the fixpoint is reached, checking
--Equation~(\ref{eq:3})-- is guaranteed to be done in a single pass
over the abstraction.  This characterization of checking ensures that
the task performed by the consumers is indeed strictly more efficient
than the certification carried out by the producers, as shown in \cite{lpar04-ai-safety-long}.

%% file: example.tex

\section{An Informal Account of our Method}
\label{sec:informal}

In this section we provide an informal account of the idea of reduced
certificate within the ACC framework by means of a very simple example.

\begin{example}\label{ex:informal}
  Consider the following program, the simple abstract domain $\tt \bot
  \sqsubseteq int \sqsubseteq real \sqsubseteq term$ that we will use
  in all our examples, and the initial calling pattern $\atom=\{{\tt
  q(X){:}\tuple{term}}\}$ which indicates that $\tt q$ can be called
  with any term as argument:
  
\begin{verbatim}
     q(X) :- p(X).            p(X) :- X = 1.0.
                              p(X) :- X = 1.
\end{verbatim} 

\noindent
A (top-down) analyzer for logic programs would start the analysis of
$\tt q(term)$ which in turn requires the analysis of $\tt p(term)$
and, as a result, the following fixpoint can be inferred: $\cert =
\{{\tt q(X){:}\tuple{term} \mapsto \tuple{real},}$ ${\tt
  p(X){:}\tuple{term} \mapsto \tuple{real}}\}$. This gives us a safe
approximation of the result of executing $\tt q(X)$. In particular, it
says that we obtain a real number as a result of executing $\tt
q(X)$. Observe that the fixpoint is sound but possibly inaccurate
since when only the second rule defining $\tt p(X)$ is executed, we
would obtain an integer number.

Given a safety policy, the next step in any approach to PCC is to
verify that $\cert$ entails such policy. For instance, if the safety
policy specifies that $ \inten = \{{\tt q(X){:}\tuple{term} \mapsto
  \tuple{term}}\}$, then clearly $\cert \sqsubseteq \inten$ holds and,
hence, $\cert$ can be used as a certificate. Similarly, a safety
policy $ \inten' = \{{\tt q(X){:}\tuple{term} \mapsto \tuple{real}}\}$
is entailed by the certificate, while $ \inten'' = \{{\tt
  q(X){:}\tuple{term} \mapsto \tuple{int}}\}$ is not.

The next important idea in ACC is that, given a valid certificate
$\cert$, a single pass of a static analyzer over it must not change
the result and, hence, this way $\cert$ can be validated. Observe that
when analyzing the second rule of $\tt p(X)$ the inferred information
$\tt X \mapsto int$ is lubbed with $\tt X \mapsto real$ which we have
in the certificate and, hence, the fixpoint does not
change. Therefore, the checker can be implemented as a non-iterating
single-pass analyzer over the certificate.  If the result of applying
the checker to $\cert$ yields a result that is different from $\cert$
an error is issued.  Once the checker has verified that $\cert$ is a
fixpoint (and thus it safely approximates the program semantics) the
only thing left is to verify that $\cert$ entails $\inten$, thus
ensuring that the validated certificate enforces the safety policy,
exactly as the certifier does.

We now turn to the key idea of reduced certificates in ACC: the
observation that \emph{any information in the certificate that the
  checker is able to reconstruct by itself in a single-pass does not
  need to be included in the certificate}. For example, if generation
of the certificate does not require iteration, then no information
needs to be included in the certificate, since by performing the same
steps as the generator the checker will not iterate. If the generator
does need to iterate, then the challenge is to find the minimal amount
of information that needs to be included in $\cert$ to avoid such
iteration in the checker.

Whether a generator requires iteration depends on the strategy used
when computing the fixpoint as well as on the domain and the program
itself (presence of loops and recursions, multivariance, etc.).  In
fact, much work has been done in order to devise optimized strategies
to reduce as much as possible iterations during analysis.  As
mentioned before,~\cite{incanal-toplas}, which will be our starting
point, presents a parametric algorithm that allows capturing a large
class of such strategies. An important observation is that whether the
checker can avoid iteration is controlled by the same factors as in
the generator, modified only by the effects of the information
included in the (reduced) certificate, that we would like to be
minimal.

As an (oversimplied) example in order to explain this idea, let us
consider two possible fixpoint strategies, each one used equally in
both the analyzer (generator) and the checker:
\begin{itemize}
\item[(1)] a strategy which first analyzes the first rule for $\tt
  p(X)$ and then the second one, and
\item[(2)] a strategy which analyzes the rules in the opposite order
  than (1).
\end{itemize}  
Assume also that the analyzer has the simple iteration rule that as
soon as an answer changes during analysis then analysis is restarted
at the top (these strategies are really too simple and no practical
analyzer would really iterate on this example, but they are useful for
illustration here --the general issue of strategies will become clear
later in the paper).

In (1), the answer $\tt X \mapsto real$ is inferred after the checking
of the first rule. Then, the second rule is analyzed which leads to
the answer $\tt X \mapsto int$ that is lubbed with the previous one
yielding $\tt X \mapsto real$. Hence, in a single pass over the
program the fixpoint is reached. Therefore, with this strategy $\tt X
\mapsto real$ can be reconstructed by the checker without iterating
and should not be included in the certificate.

However, with strategy (2) we first obtain the answer $\tt X \mapsto
int$. Then, after the analysis of the first rule, $\tt X \mapsto real$
is inferred.  When lubbing it with the previous value, $\tt X \mapsto
int$, the answer obtained is $\tt X \mapsto real$.  Since the answer
has changed the analyzer starts a new iteration in which it reanalyzes
the second rule with the new answer $\tt X \mapsto real$. Since now
nothing changes in this iteration the fixpoint is reached.

The key idea is that, if strategy (2) is used, then more than one
iteration is needed to reach the fixpoint. Hence the certificate
cannot be empty and instead it has to include (some of) the analysis
information.  The conclusion is that the notion of reduced certificate
is strongly related to the strategy used during analysis and
checking. \hfill $\Box$
\end{example}

The remainder of the article will formalize and discuss in detail each
of the above steps and issues.

%% file: analysis.tex

\section{Generation of Certificates in Abstraction-Carrying Code}\label{sec:an-abstr-interpr}

This section recalls ACC and the notion of full certificate in the
context of (C)LP \cite{lpar04-ai-safety-long}.  This programming
paradigm offers a good number of advantages for ACC, an important one
being the maturity and sophistication of the analysis tools available
for it.  It is also a non-trivial case in many ways, including the
fact that logic variables and incomplete data structures essentially
represent respectively pointers and structures containing pointers
(see also the arguments and pointers to literature in
Section~\ref{sec:introduction} which provide evidence that our
approach is applicable essentially directly to programs in other
programming paradigms, including their bytecode representations).

Very
briefly, {\em terms} are constructed from variables $x \in \vars$, {\em
  functors} (e.g., $f$) and {\em predicates} (e.g., $p$). We denote by
$\{x_1 / t_1, \ldots, x_n / t_n\}$ the {\em substitution} $\sigma$,
where $x_i \neq x_j$, if $i \neq j$, and $t_i$ are terms. A {\em
  renaming} is a substitution $\rho$ for which there exists the
inverse $\rho^{-1}$ such that $\rho \rho^{-1} \equiv \rho^{-1} \rho
\equiv {\it id}$. A {\em constraint} is a conjunction of expressions
built from predefined predicates (such as inequalities over the reals)
whose arguments are constructed using predefined functions (such as
real addition). An {\em atom} has the form $p(t_1,...,t_n)$ where $p$
is a predicate symbol and $t_i$ are terms.  A {\em literal} is either
an atom or a constraint.  A {\em rule} is of the form $H \mbox{\tt :-}
D$ where $H$, the {\em head}, is an atom and $D$, the {\em body}, is a
possibly empty finite sequence of literals.  A {\em constraint logic
  program} $P \in {\it Prog}$, or {\em program}, is a finite set of
rules.  Program rules are assumed to be normalized: only distinct
variables are allowed to occur as arguments to atoms. Furthermore, we
require that each rule defining a predicate $p$ has identical sequence
of variables $x_{p_1}, \ldots x_{p_n}$ in the head atom, i.e.,
$p(x_{p_1}, \ldots x_{p_n})$.  We call this the {\em base form} of
$p$. This is not restrictive since programs can always be normalized.

\subsection{The Analysis Algorithm}

\begin{algorithm}[th]
\caption{Generic Analyzer for Abstraction-Carrying Code}

\label{algo:gen-anal}\begin{algorithmic} [1]
\renewcommand{\baselinestretch}{0.95} 
\small

\Statex {\sf Initialization of global data structures:} ${\it DAT{=}AT{=}\emptyset}$

\Function{\analyzerA}{$\atom \subseteq \aatom,\strategy \in \qhs$}

\For{$A:\CP{} \in \atom$}
\State {\sf add\_event}(${\it newcall}(A:\CP{}),\strategy$);
\EndFor

\While{$E{=} {\sf next\_event}(\strategy)$} \label{event}
\If {$E {=} {\it newcall}(A:\CP{})$}
  {\sc new\_call\_pattern}($A:\CP{},\strategy$); \label{newcall}
\ElsIf {$E {=} {\it updated}(A:\CP{})$}
 {\sc add\_dependent\_rules}($A:\CP{},\strategy$);\label{update}
\ElsIf {$E {=} {\it arc}(R)$}
  {\sc process\_arc}($R,\strategy$); \label{arc}
\EndIf
\EndWhile
\State {\bf return} {\it AT};
\EndFunction

\Statex
\Procedure{new\_call\_pattern}{$A:\CP{} \in \aatom,\strategy \in \qhs$}
\ForAll{rule $A_k :- B_{k,1}, \ldots, B_{k,n_k}$}
\State $\CP{}_0$ :{=}{\sf Aextend}$(\CP{}, vars(\ldots,B_{k,i}, \ldots))$;
\State $\CP{}_1$ :{=} {\sf Arestrict}$(\CP{}_0, vars(B_{k,1}))$;
\State {\sf add\_event}($arc(A_k:\CP{} \Rightarrow{} [\CP{}_0] ~B_{k,1}:\CP{}_1$),$\strategy$);\label{12}
\EndFor
\State {\sf add\_answer\_table}($A:\CP{} {\mapsto} \bot$); \label{false}

\EndProcedure
\Statex
\Procedure{process\_arc}{$H_k:\CP{}_0 \Rightarrow  [\CP{}_1]
  ~B_{k,i}:\CP{}_2 \in \depend$, $\strategy \in \qhs$}
\If {$B_{k,i}$ is not a constraint}  \label{20}
\State add $H_k:\CP{}_0 \Rightarrow  [\CP{}_1] ~B_{k,i}:\CP{}_2$ to
$\it DAT$; \label{21}
\EndIf
\State $W$ :{=} $vars(H_k,B_{k,1}, \ldots, B_{k,n_k})$; \label{Wcheck}
\State $\CP{}_3$ :{=} {\sc get\_answer}($B_{k,i}:\CP{}_2, \CP{}_1, W,\strategy$); \label{23}
\If {$\CP{}_3 \neq \bot$ and $i \neq n_k$}\label{notlast} \label{24}
\State $\CP{}_4$ :{=} {\sf Arestrict}$(\CP{}_3, vars(B_{k,i+1}))$;
\State {\sf add\_event}( $arc(H_k:\CP{}_0 \Rightarrow [\CP{}_3]
~B_{k,i+1}:\CP{}_4$),$\strategy$); \label{add-event-1}
\ElsIf {$\CP{}_3 \neq \bot$ and $i {=} n_k$}\label{last}
 \State $\AP{}_1$ :{=} {\sf Arestrict}$(\CP{}_3, vars(H_k))$;
{\sc insert\_answer\_info}($H:\CP{}_0 {\mapsto} \AP{}_1,\strategy$);  \label{relast}
\EndIf
\EndProcedure

\Statex
\Function{get\_answer}{$L:\CP{}_2 \in \aatom,\CP{}_1 \in \dom,W
  \subseteq \vars,\strategy \in \qhs$}
\If{$L$ is a constraint}
 {\bf return} {\sf Aadd}$(L,\CP{}_1)$; \label{xx}
\Else  $~\AP{}_0$ :{=} {\sc lookup\_answer}$(L:\CP{}_2,\strategy)$;  \label{32}
       $\AP{}_1$ :{=} {\sf Aextend}$(\AP{}_0,W)$; \label{33}
\State ~~{\bf return} {\sf Aconj}$(\CP{}_1,\AP{}_1)$; \label{34}
\EndIf
\EndFunction
\Statex

\Function{lookup\_answer}{$A:\CP{} \in \aatom,\strategy \in \qhs$}
\If{there exists a renaming $\sigma$ s.t. $\sigma(A:\CP{}) {\mapsto}
  \AP{}$ in {\it AT}} \label{36}
\State {\bf return} $\sigma^{-1}(\AP{})$; \label{36-1}
\Else  {\sf ~add\_event}(${\it newcall}(\sigma(A:\CP{})),\strategy$) where $\sigma$ is
renaming s.t.\   $ \label{38}
\sigma(A)$ in base form;
\State {\bf return} $\bot$;
\EndIf
\EndFunction
\Statex

\Procedure{insert\_answer\_info}{$H:\CP{} {\mapsto} \AP{} \in \entries,\strategy
  \in \qhs$}
\State $\AP{}_0$ :{=} {\sc lookup\_answer}$(H:\CP{})$; $\AP{}_1$ :{=} {\sf Alub}$(\AP{},\AP{}_0)$;\label{41}
\If{$\AP{}_0 \neq \AP{}_1$}\label{411}
\State  {\sf add\_answer\_table}($(H:\CP{} {\mapsto} \AP{}_1)$; \label{43}

\State {\sf add\_event}(${\it updated}(H:\CP{}),\strategy$);  \label{45}
\EndIf
\EndProcedure
\Statex

\Procedure{add\_dependent\_rules}{$A:\CP{} \in \aatom,\strategy \in \qhs$}
\ForAll{arc of the form $H_k:\CP{}_0 \Rightarrow [\CP{}_1]
  ~B_{k,i}:\CP{}_2$ in graph {\bf where} there exists renaming $\sigma$
  s.t.  $A:\CP{} {=} (B_{k,i}:\CP{}_2)\sigma$} \label{37}
\State {\sf add\_event}($arc(H_k:\CP{}_0 \Rightarrow [\CP{}_1] ~B_{k,i}:\CP{}_2),\strategy$); \label{55}
\EndFor
\EndProcedure
\renewcommand{\baselinestretch}{0.7}
\end{algorithmic}
\end{algorithm}

\bigskip

Algorithm \ref{algo:gen-anal} has been presented in
\cite{incanal-toplas} as a generic description of a fixpoint
algorithm which generalizes those used in state-of-the-art analysis engines,
such as the one in \ciaopp~\cite{ciaopp-sas03-journal-scp},
PLAI \cite{ai-jlp,anconsall-acm}, GAIA
  \cite{LeCharlier94:toplas}, and the CLP($\cal R$) analyzer
  \cite{softpe}. It has the description domain $D_\alpha$ (and functions on
  this domain) as parameters.  Different domains give analyzers which
  provide different kinds of information and degrees of accuracy.  
In order to analyze a program, traditional (goal dependent) abstract
interpreters for (C)LP programs receive as input, in addition to the
program $P$ and the abstract domain $\dom$, a set $\atom \subseteq \aatom$
of Abstract Atoms (or {\em call patterns}).
Such call patterns are pairs of the form $A:{\it CP}$ where $A$ is a
procedure descriptor and ${\it CP}$ is an abstract substitution (i.e.,
a condition of the run-time bindings) of $A$ expressed as ${\it CP}
\in \dom$. For brevity, we sometimes omit the subscript $\alpha$ in
the algorithms. The analyzer of Algorithm \ref{algo:gen-anal},
\analyzerA, constructs an \emph{and--or graph}~\cite{bruy91} (or
analysis graph) for $\atom$ which is an abstraction of the (possibly
infinite) set of (possibly infinite) execution paths (and-or trees)
explored by the concrete execution of the initial calls described by
$\atom$ in $P$.  Let $S_P^\alpha$ be the abstract semantics of the
program for the call patterns $\atom$ defined in \cite{bruy91}.
Following the notation in Section~\ref{sec:basics-abstr-carry}, the
analysis graph --denoted as $\p_\alpha$-- corresponds to (or safely
approximates) {\rm lfp}($S_P^\alpha$).

The program analysis graph is implicitly represented in the algorithm
by means of two global data structures, the {\em answer table} $\it AT$  and the
{\em dependency arc table} $\it DAT$, both initially empty as shown at
the beginning of Algorithm~\ref{algo:gen-anal}.\footnote{Given the information 
in these, it is straightforward to
  construct the graph and the associated program-point annotations.}

\begin{definition}[answer and dependency arc table]
Let $P \in {\it Prog}$ be a program  and $\dom$ an abstract domain.

\begin{itemize}
\item An {\em Answer Table} (${\it AT} \subseteq \entries$) for $P$ and $\dom$ is 
 a set of entries of the form  $A:\CP{} {\mapsto} \AP{} \in \entries$ 
  where $A:\CP \in \aatom$, $A$ is always in  base form 
and $\CP$ and $\AP$ are abstract substitutions in $\dom$. 

\item {\em A Dependency Arc Table} (${\it DAT} \subseteq \depend$) for $P$ and $\dom$ 
  is a set of {\em
  dependencies} of the form   
  $A_k:{\it CP_0} \Rightarrow [\CP_1]~B_{k,i}:{\it CP}_2 \in \depend$, where $A_k \mbox{ :- }
  B_{k,1},\ldots,B_{k,n}$ is a program rule in $P$ and
  $\CP_0,\CP_1,\CP_2$ are abstract substitutions in $\dom$.

\end{itemize}
\end{definition}

Informally, an entry $A:{\it CP} \mapsto {\it AP}$ in ${\it AT}$
 should be interpreted as ``the answer
  pattern for calls to $A$ satisfying precondition (or call pattern)
  $\CP{}$ meets postcondition (or answer pattern), $\AP{}$.''
 Dependencies are used for efficiency. As we will explain later, 
 Algorithm \ref{algo:gen-anal} finishes when there are no more events to be
 processed (function \analyzerA). This happens when the answer 
 table $\it AT$ reaches a fixpoint. Any entry $\it A:CP \mapsto AP$
 in $\it AT$ is generated by analyzing all rules associated
 to $A$ (procedure {\sc new\_call\_pattern}).  Thus, if we have a rule of the form 
  $A_k \mbox{ :- } B_{k,1},\ldots,B_{k,n}$, we know that the answer
  for $A$ depends on the answers for all literals in the body of the
  rule.  We annotate this fact in $\it DAT$ by means of the dependencies 
 $A_k:{\it CP} \Rightarrow
  [\CP_{k,{i-1}}]~B_{k,i}:{\it CP}_{k,i}$, $i \in \{1,..n\}$,  which
  mean that the answer for $A_k:{\it CP}$
  depends on the answer for $B_{k,i}:{\it CP}_{k,i}$, also stored in
  $\it AT$. Then if during the analysis, the answer for
  $B_{k,i}:{\it CP}_{k,i}$ changes,  the \emph{arc} 
  $A_k:{\it CP} \Rightarrow [{\it CP}_{k,i-1}] B_{k,i}:{\it CP}_{k,i}$ must be
  reprocessed in order to compute the ``possibly'' 
  new answer for $A_k:{\it CP}$. This is
  to say that the rule for $A_k$ has to be processed again starting
  from atom $B_{k,i}$. Thus, as we will see later, dependency arcs are
  used for forcing recomputation until a fixpoint is reached.
The remaining part $\CP{}_{k,i-1}$ is the
program annotation just before $B_{k,i}$ is reached and contains
information about all variables in rule $k$.
 $\CP{}_{k,i-1}$ is not really necessary, 
but is included for
efficiency.

Intuitively, the analysis algorithm is a graph traversal algorithm
which places entries in the answer table $\it AT$ and dependency arc
table $\it DAT$ as
new
 nodes and arcs in the program analysis graph are encountered.  To
capture the different graph traversal strategies used in different
fixpoint algorithms,  a {\em prioritized event
  queue} is used.
We use $\strategy \in \qhs$ to refer to a
\emph{Queue Handling Strategy} which a particular instance of the
generic algorithm may use. 
Events are
of three forms:

\begin{itemize}
\item ${\it newcall}(A:{\it CP}{})$ which indicates that a new call pattern
for literal $A$ with abstract substitution $\CP{}$ has been encountered.
\item $arc(H_k:\_ \Rightarrow [\ \_ \ ] ~B_{k,i}:\_)$ which indicates that
  the rule with $H_k$ as head needs to be (re)computed from the
  position $k,i$.
\item ${\it updated}(A:{\it CP}{})$ which indicates that the answer  to
call pattern $A$ with abstract substitution $\CP{}$ has been changed
in $\it AT$.
\end{itemize}

\noindent
The algorithm is defined in terms of five abstract operations on the
domain $D_{\alpha}$:

\begin{itemize}
\item ${\sf Arestrict(}\CP{},V{\sf )}$ performs the abstract
  restriction of an abstract substitution $\CP{}$ to the set of variables in the set $V$.
\item ${\sf Aextend(}\CP{},V{\sf )}$ extends the abstract substitution $\CP{}$ to the
  variables in the set $V$.
 \item ${\sf Aadd}(C,\CP{})$ performs the abstract operation of
   conjoining the actual constraint $C$ with the abstract substitution $\CP{}$.
\item $\sf Aconj(\CP{}_1,\CP{}_2)$ performs the abstract conjunction
  of two abstract substitutions.
\item $\sf Alub(\CP{}_1,\CP{}_2)$ performs the abstract disjunction of
  two  abstract substitutions.
\end{itemize}
Apart from the parametric 
domain-dependent functions, the algorithm has several other undefined
functions. The functions {\sf add\_event} and {\sf next\_event}
respectively push an event to the priority queue and pop
the event of highest priority, according to $\strategy$.
When an arc $H_k:\CP{} \Rightarrow [\CP{}'']
\ B_{k,i}:\CP{}'$ is added to $\it DAT$, it replaces any
other arc of the form $H_k:\CP{} \Rightarrow [\ \_\ ]\  B_{k,i}:\_$
(modulo renaming) in the table and the priority queue.
Similarly when an entry $H_k:\CP{} \mapsto \AP{}$ is added to the $\it
AT$ ({\sf add\_answer\_table}), it replaces any entry of the form $H_k:\CP{} \mapsto \_$
(modulo renaming).
Note that the underscore ($\_$) matches any description, and that there is
at most one matching entry in $\it DAT$  or $\it AT$ at any time.

More details on the algorithm
can be found in \cite{incanal-toplas,inc-fixp-sas-semi-short}.
Let us briefly explain
its main procedures:

\begin{itemize}
\item The algorithm centers around the processing of
events on the priority queue, which repeatedly removes the highest
priority event (Line \ref{event}) and calls the appropriate
event-handling function (L\ref{newcall}-\ref{arc}).
\item The function {\sc
  new}{\sc \_call}{\sc \_}{\sc pattern} initiates processing of all the rules for the
definition of the internal literal $A$, by adding arc events for each
of the first literals of these rules (L\ref{12}).
Initially, the answer for the call pattern is set to $\bot$
(L\ref{false}).

\item The procedure {\sc process\_arc} performs the core of the analysis.  It
performs a single step of the left-to-right traversal of a rule body.

\begin{itemize}
\item If the literal $B_{k,i}$ is not a constraint (L\ref{20}), the
  arc is added to $\it DAT$ (L\ref{21}).
\item Atoms are processed by function {\sc get\_answer}:

\begin{itemize}
\item   Constraints are
simply added to the current description (L\ref{xx}). 

\item In the case of
literals, 
the function {\sc lookup\_answer} first looks up an answer for the
given call pattern in $\it AT$ (L\ref{36}) and if it is not
found, it places a \emph{newcall} event (L\ref{38}).  When it finds
one, then this answer is extended to the variables in the rule the
literal occurs in (L\ref{33}) and \emph{conjoined} with the current
abstract substitution (L\ref{34}).
The resulting answer (L\ref{23}) is either used to generate a new arc
event to process the next literal in the rule, if $B_{k,i}$ is not the
last one (L\ref{24}); otherwise, the new answer is computed by {\sc
  insert\_answer\_info}.  
\end{itemize}
\end{itemize}

\item The part of the algorithm that is more relevant to the
  generation of 
  reduced certificates is within {\sc insert\_answer\_info}. The new
  answer for the rule is \emph{combined} with the current answer in
  the table (L\ref{41}). If the fixpoint for such call has not been
  reached, then the corresponding entry in $\it AT$ is updated
  with the combined answer (L\ref{43}) and an updated event is added
  to the queue (L\ref{45}).

\item The purpose of an updated event is that the function {\sc
    add\_dependent\_rules} (re)processes those calls which depend on
  the call pattern $A:\CP{}$ whose answer has been updated
  (L\ref{37}).  This effect is achieved by adding the arc events for
  each of its dependencies (L\ref{55}).  The fact that dependency arcs
  contain information at the level of body literals, identified by a
  pair $k,i$, allows reprocessing only those rules for the predicate
  which depend on the updated pattern. Furthermore, those rules are
  reprocessed precisely from the body atom whose answer has been
  updated. If, instead, dependencies were kept at the level of rules,
  rules would need to be reprocessed always from the leftmost
  atom. Furthermore, if dependencies were kept at the level of
  predicates, all rules for a predicate would have to be reprocessed
  from the leftmost atom as soon as an answer pattern it depended on
  were updated. 
\end{itemize}
In the following section, we illustrate the algorithm by means of an
example.

\subsection{Running Example}

Our running example is the program {\tt rectoy} taken from
\cite{RR98}. We will use it to illustrate our algorithms and show that
our approach improves on state-of-the-art techniques for reducing the
size of certificates.
Our approach can deal with the very wide class of properties
for which abstract interpretation has been proved useful (for
example in the context of LP this includes variable sharing, determinacy,
non-failure, termination, term size, etc.). For brevity and concreteness,
in all our examples abstract substitutions simply assign an abstract
value in the simple domain introduced in Section~\ref{sec:informal} to
each variable in a set $V$ over which each such substitution ranges.
We use {\tt term} as the most general type (i.e., $\tt term$
corresponds to all possible terms).
For brevity, variables whose regular type is $\tt term$ are often not
shown in abstract substitutions. Also, when it is clear from the
context, an abstract substitution for an atom $p(x_1, \ldots, x_n)$ is shown as a tuple
$\tuple{t_1,\ldots,t_n}$, such that each value $t_i$ indicates the
type of $x_i$.
The most general substitution $\top$ assigns \texttt{term} to all
variables in $V$.  The least general substitution $\bot$ assigns the
empty set of values to each variable.

\begin{example}\label{ex:rectoy1}
  Consider the \ciao\ version of procedure {\tt rectoy}
  \cite{RR98} and the call pattern $\tt rectoy(N,M):\tuple{int,term}$ which
  indicates that external calls to {\tt rectoy} are performed with an
  integer value, \texttt{int}, in the first argument \texttt{N}:

\[ \begin{array}{llll}
\texttt{rectoy(N,M)}&\texttt{:-} &\texttt{N~{=}~0, M~{=}~0.} \\
\texttt{rectoy(N,M)}&\texttt{:-} & \texttt{N1 is N-1, rectoy(N1,R), M is
  N1+R.}
\end{array}\]

\noindent
We now briefly describe four main steps carried out in the analysis using some 
  $\strategy \in \qhs$:

\begin{itemize}
\item[A.]  The initial event ${\it newcall}({\tt rectoy(N,M):\tuple{int,term}})$
  introduces the arcs $A_{1,1}$ and $A_{2,1}$ in the queue, each one
  corresponds to the rules in the order above:
\[
\begin{array}{l}
A_{1,1} \equiv {\it arc}({\tt rectoy(N,M):\tuple{int,term} \Rightarrow{} [\{N/int\}] ~
  N{=}0:\{N/int\}}) \\
A_{2,1}\equiv {\it arc}({\tt rectoy(N,M):\tuple{int,term} \Rightarrow{} [\{N/int\}] ~ N1~ is~
  N-1:\{N/int\}})
\end{array}
\]
The initial answer $E_1 \equiv \tt
  rectoy(N,M):\tuple{int,term} \mapsto \bot$
 is inserted in $\it AT$.

\item[B.] Assume that $\strategy$ assigned higher priority to
  $A_{1,1}$.  The procedure {\sc get\_answer} simply adds the
  constraint $\tt N{=}0$ to the abstract substitution $\tt \{N/int\}$. Upon
  return, as it is not the last body atom, the following arc event is
  generated:
\[
\begin{array}{l}
A_{1,2} \equiv {\it arc}({\tt rectoy(N,M):\tuple{int,term} \Rightarrow{} [\{N/int\}] ~
  M{=}0:\{M/term\}})
\end{array}
\]

Arc $A_{1,2}$ is handled exactly as $A_{1,1}$ and {\sc get\_answer}
simply adds the constraint $\tt M{=}0$, returning $\tt \{N/int,M/int\}$.
As it is the last atom in the body (L\ref{last}), procedure {\sc
insert\_answer\_info} computes $\alub$ between $\bot$ and the above
answer and overwrites $E_1$ with:

\[ \fbox{$E'_1 \equiv
   {\tt rectoy(N,M):\tuple{int,term} \mapsto \tuple{int,int}}$} \]

\noindent
Therefore, the event 
  $U_1 \equiv  {\it updated}({\tt rectoy(N,M):\tuple{int,term}})$ is
introduced in the queue. Note that no dependency has been originated
during the processing of this rule (as both body atoms are
constraints).

\item[C.] Now, $\strategy$ can choose between the processing of $U_1$ or
  $A_{2,1}$. Let us assume that $A_{2,1}$ has higher priority.  For
  its processing, we have to assume that predefined functions ``$\tt
  -$'', ``$\tt +$'' and ``$\tt is$'' are dealt by the algorithm as
  standard constraints by just using the following information
  provided by the system:

\[
\begin{array}{lll} 
E_2 &\equiv& \tt C ~is~ A+B: ~\tuple{int,int,term}  \mapsto
\tuple{int,int,int} \\  
E_3 &\equiv&  \tt C ~is~ A-B: ~\tuple{int,int,term}  \mapsto
\tuple{int,int,int} \\  
\end{array}
\]
where the three values in the abstract substitutions correspond to
variables $\tt A$, $\tt B$, and $\tt C$, in this order.
In particular, after analyzing the subtraction with the initial call
pattern, we infer that $\tt N1$ is of type $\tt int$ and no dependency
is asserted.  Next, the arc:
\[
\begin{array}{lll}
A_{2,2}& \equiv & {\it arc}({\tt rectoy(N,M):\tuple{int,term} \Rightarrow{}} \\
& & {\tt [\{N/int,N1/int\}] ~ rectoy(N1,R):\tuple{int,term}})
\end{array}
\]
is introduced in the queue and the corresponding dependency is stored
in $\it DAT$. The call to {\sc get\_answer} returns the
current answer $E'_1$. 
Then, we use this answer as call pattern to
process the last addition by creating a new arc $A_{2,3}$.

\[\begin{array}{lll}
A_{2,3} & \equiv & {\it arc}({\tt rectoy(N,M):\tuple{int,term} \Rightarrow{}}\\
&& {\tt [\{N/int,N1/int,R/int\}] ~ M ~is~ N1+R:\{N1/int,R/int\}})
\end{array}\]
Clearly, the processing of $A_{2,3}$ does not change the final answer
$E'_1$.  Hence, no more updates are introduced in the queue.

\item[D.] Finally, we have to process the event $U_1$ introduced in
  step B to which $\Omega$ has assigned lowest priority. The procedure
  {\sc add\_dependent\_rules} finds the dependency corresponding to
  arc $A_{2,2}$ and inserts it in the queue.  This relaunches an arc
  identical to $A_{2,2}$. This in turn launches an arc identical to
  $A_{2,3}$. However, the reprocessing does not change the fixpoint
  result $E'_1$ and the analysis terminates computing as answer table the entry 
  $E'_1$ and as unique dependency arc $A_{2,2}$.
\end{itemize}

\noindent
Figure~\ref{fig-dep} shows the analysis graph for the analysis above.
The graph has two sorts of nodes. Those which correspond to atoms are
called ``OR-nodes.'' An OR-node of the form $^{{\it CP}}A^{AP}$ is
interpreted as: the answer for the call pattern $A:\CP{}$ is $\AP{}$.
For instance, the OR-node

\[ {\tt
   ^{\{N1/int\}}rectoy(N1,R)^{\{N1/int,R/int\}}}\]
\noindent
indicates that, when the atom $\tt rectoy(N1,R)$ is called with
the abstract substitution $\tt \tuple{int,term}$, the answer computed is $\tt
\tuple{int,int}$.
As mentioned before, variables whose type is {\tt term} will often not
be shown in what follows.
Those nodes which correspond to rules are called ``AND-nodes.'' In 
Figure \ref{fig-dep}, they appear within a dotted box and contain the head of
the corresponding clause. Each AND-node has as children as many
OR-nodes as there are atoms in the body.  If a child OR-node is
already in the tree, it is not expanded any further and the currently
available answer is used.
For instance, the analysis graph in the figure at hand contains two
occurrences of the abstract atom $\tt rectoy(N,M):\tuple{int,term}$
(modulo renaming), but only one of them (the root) has been expanded.
This is depicted by a dashed arrow from the non-expanded occurrence to
the expanded one.

\begin{center}
 \begin{figure}[t]
\small \begin{center} 
\fbox{
\begin{minipage}{12cm}\centering
  $\xymatrix@!R=0.0pt@!C=45pt{
    & & E_2: ^{0}{\tt rectoy(N,M)^{11}}\ar[dl]^{A_{1}}\ar[dr]^{A_{2}} & & &\\
   &*[F.]{\tt rectoy(N,M)}\ar[dl]^{A_{1,1}}\ar[d]^{A_{1,2}} & &
   *[F.]{{\tt rectoy(N,M)}}\ar[dl]^{A_{2,1}}\ar[d]^{A_{2,2}}\ar[dr]^{A_{2,3}} & & \\
 ^{1}{\tt N=0}^{2} & ^{3}{\tt M=0}^{4}&^{5}{\tt N1~is~ N-1}^{6}& ^{7}{\tt rectoy(N1,R)}^{8}\ar@{-->}[uul] & ^{9}{\tt M~
    is~ N1-R}^{10}
 \\}$

$\begin{array}{l@{~:~}l}
0,1,2,3 & \tt \{N/int\}\\
4 & \tt \{N/int,M/int\}\\
5 & \tt  \{N/int\} \\
6 & \tt \{N/int,N1/int\}
\end{array}
$
$\begin{array}{l@{~:~}l}
7 & \tt \{N1/int\}\\
8 & \tt \{N1/int,R/int\}\\
9 &\tt  \{M/int,R/int,N1/int\} \\
10 & \tt \{M/int,R/int,N1/int\} \\
11 & \tt \{N/int, M/int\}
\end{array}
$

\end{minipage}
}\end{center}\caption{Analysis Graph for our Running Example}
\label{fig-dep}
\end{figure}
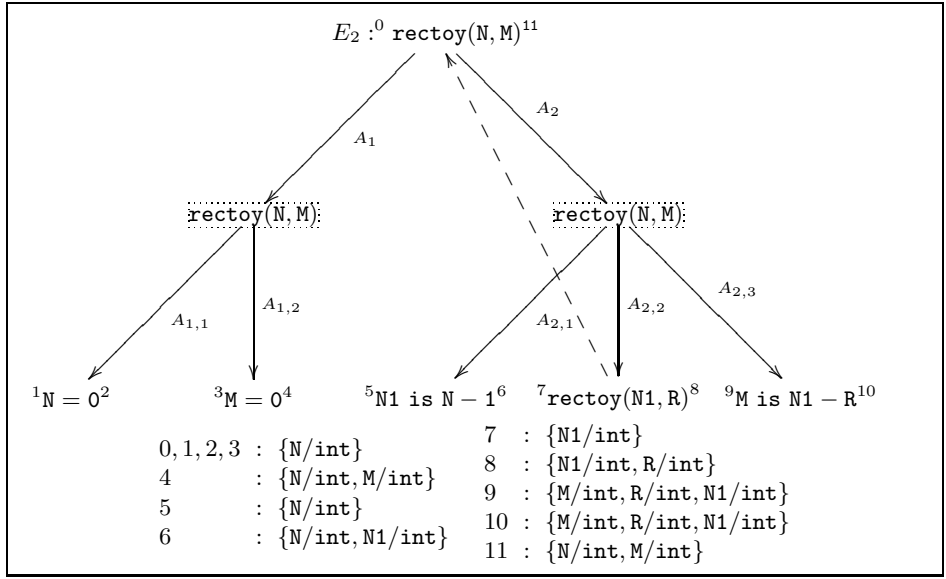
\end{center}

\noindent
The answer table $\it AT$ contains entries for the different OR-nodes which
appear in the graph. In our example $\it AT$ contains $E'_1$
associated to the (root) OR-node discussed above.
Dependencies in $\it DAT$ indicate direct relations
among OR-nodes. An OR-node $A_F:{\it CP}_F$ \emph{depends on} another
OR-node $A_T:{\it CP}_T$ iff the OR-node $A_T:{\it CP}_T$ appears in the body of
some clause for $A_F:{\it CP}_F$.
For instance, the dependency $A_{2,2}$
indicates that the OR-node $\tt rectoy(N1,R): \tuple{int,term}$ is
 used in the OR-node $\tt rectoy(N,M): \tuple{int,term} $.  Thus,
if the answer pattern for $\tt rectoy(N1,R): \tuple{int,term}$ is ever
updated, then we must reprocess the OR-node $\tt rectoy(N,M):
\tuple{int,term}$. \hfill $\Box$

\end{example}

\subsection{Full Certificate}

The following definition corresponds to the essential idea in the ACC
framework --Equations~(\ref{eq:1}) and (\ref{eq:2})-- of using a
static analyzer to generate the certificates. The analyzer corresponds
to Algorithm~\ref{algo:gen-anal} and the certificate is the
\emph{full} answer table.

\begin{definition}[full certificate]\label{def:gen-cert}
  We define function \certifierA~: $\prog \times \adom \times 2^{\aatom}
  \times \aint \times \qhs \mapsto \acert$ which takes  $P \in Prog$,
  $\dom \in \adom$,  $\atom \subseteq
  \aatom$, $\inten \in \aint$, $\strategy \in \qhs$ and returns as
  \emph{full certificate}, $\certificateA \in \acert$, the answer table
  computed by \analyzerA$(\atom,\strategy)$ for $P$ in $\dom$ iff $\certificateA\sqsubseteq \inten$.
\end{definition}
If the inclusion does not hold, we do not have a certificate. This can
happen either because the program does not satisfy the policy or
because the analyzer is not precise enough. In the latter case, a
solution is to try analyzing with a more precise (and generally more
expensive) abstract domain. In the former case (the program does not
satisfy the policy), this can be due to two possible reasons. A first
one is that we have formalized a policy which is unnecessarily
restrictive, in which case the solution is to weaken it. The other
possible reason is that the program actually violates the policy,
either inadvertently or on purpose. In such a case there is of course no
way a certificate can be found for such program and policy.

\begin{example}\label{ex:rectoy2} Consider the safety policy
  expressed by the following specification
  $\inten: {\tt
  rectoy(N,M):}$ ${\tt \tuple{int,term} {\mapsto} \tuple{int,real}}$.  The
  certifier in Definition~\ref{def:gen-cert} returns as valid certificate
  the single entry $E'_1$. Clearly $E'_1 \sqsubseteq \inten$ since
  $\tt \bot \sqsubseteq int \sqsubseteq real \sqsubseteq term$.  \hfill $\Box$
\end{example}

%% file: redundant.tex

\section{Abstraction-Carrying Code with Reduced Certificates}
\label{sec:reduced-certificates}

As already mentioned in Section~\ref{sec:introduction}, in the ACC
framework, since this certificate contains the fixpoint, a
\emph{single pass} over the analysis graph is sufficient to validate
such certificate on the consumer side.
The key observation in order to reduce the size of certificates within
the ACC framework is that certain entries in a certificate may be
\emph{irrelevant}, in the sense that the checker is able to reproduce
them by itself in a single pass.  The notion of \emph{relevance} is
directly related to the idea of recomputation in the program analysis
graph. Intuitively, given an entry in the answer table ${\it
  A:CP \mapsto \it AP}$, its fixpoint may have been computed in
several iterations from $\bot, ~ {\it AP}_0,~ {\it AP}_1,\ldots$ until
${\it AP}$. For each change in the answer, an event ${\it
  updated}$(${\it A:CP}$) is generated during the analysis.  The
above entry is \emph{relevant} in a certificate (under some strategy)
when its updates launch the recomputation of other arcs in the graph
which \emph{depend} on $A:{\it CP}$ (i.e., there is a dependency
from it in the table).  Thus, unless ${\it A:CP\mapsto AP}$ is
included in the (reduced) certificate, a \emph{single-pass} checker
which uses the same strategy as the code producer will not be able to
validate the certificate. Section~\ref{sec:ident-redund-inform} 
identifies redundant updates which should not be considered.
In Section \ref{sec:noti-reduc-cert}, we
characterize formally the notion of reduced certificate containing
only relevant answers. Then, in Section~\ref{sec:gener-cert-with}, we
instrument an analysis algorithm to identify relevant answers and
define a certifier based on the instrumented analyzer which generates
reduced certificates.

\subsection{Identifying Redundant Updates}\label{sec:ident-redund-inform}

According to the above intuition, we are interested in determining
when an entry in the answer table has been ``updated'' during the
analysis and such changes affect other entries. There is a special
kind of updated events  which can be directly considered
irrelevant and correspond to those updates which launch a 
redundant computation
(like the $U_1$ event generated in step B of Example~\ref{ex:rectoy1}). We
write ${\it DAT|_{{\it A:CP}}}$ to denote  the set of arcs of the form ${\it H:CP_0
  \Rightarrow [CP_1] \ B:CP_2} \in \depend$ in the current dependency arc table 
which depend on $A:{\it CP}$, i.e., such
that ${\it A:CP = (B:CP_2)\sigma}$ for
some renaming $\sigma$.

\begin{definition}[redundant update]\label{def:redundant}
  Let $P \in {\it Prog}$, $\atom \subseteq \aatom$ and $\Omega \in \qhs$.  We say
  that an event ${\it updated}$(${\it A:CP}$) which appears in the
  prioritized event
  queue during the analysis of $P$ for $\atom$ is \emph{redundant}
  w.r.t.\ $\Omega$ if, when it is generated, ${\it DAT|_{{\it
        A:CP}}=\emptyset}$.
\end{definition}

\noindent
It should be noted that redundant updates can only be generated by
updated events for call patterns which belong to $\atom$, i.e., to the
initial set of call patterns. Otherwise, ${\it DAT|_{{\it A:CP}}}$
cannot be empty.  Let us explain the intuition of this. The reason is
that whenever an event ${\it updated}({\it A:CP})$, ${\it A:CP} \not
\in \atom$, is generated is because a rule for $A$ has been completely
analyzed. Hence, a corresponding call to {\sc insert\_answer\_info}
for ${\it A:CP}$ (L\ref{relast} in Algorithm~\ref{algo:gen-anal}) has
been done.  If such a rule has been completely analyzed then all
its arcs were introduced in the prioritized event queue. Observe that
the first time that an arc is introduced in the queue is because a
call to procedure {\sc new\_call\_pattern} for ${\it A:CP}$ occurred,
i.e., a ${\it newcall}(A:CP)$ event was analyzed.  Consider the first
event {\it newcall} for ${\it A:CP}$. If ${\it A:CP} \not \in \atom$,
then this event originates from the analysis of some other arc of the
form $H:{\it CP_0} \Rightarrow [{\it CP}_1] {\it A:CP}$ for which
${\it A:CP}$ has no entry in the answer table.  Thus, the dependency
$H:{\it CP_0} \Rightarrow [{\it CP_1}] {\it A:CP}$ was added to ${\it
  DAT}$. Since dependencies are never removed from ${\it DAT}$, then
any later updated event for ${\it A:CP}$ occurs under the condition
${\it DAT|_{{\it A:CP}}\not =\emptyset}$.  Even if it is possible to
fix the strategy and define an analysis algorithm which does not
introduce redundant updates, we prefer to follow as much as possible
the generic one.

\begin{example}
In our running example $U_1$ is redundant for $\Omega$ at the
moment it is generated. However, since the event has been given low
priority its processing is delayed until the end and, in the
meantime, a dependency from it has been added.
This causes the unnecessary
redundant recomputation of the second arc $A_{2,2}$ for $\tt rectoy$.
\hfill $\Box$
\end{example}

\noindent
Note that redundant updates are indeed events which if processed
immediately correspond to ``nops''.

The following proposition ensures the correctness of using a queue
handling strategy which assigns the highest priority to redundant
updates. This result can be found in \cite{incanal-toplas}, where it is
stated that \analyzerA~is correct independently of the order in which
events in the prioritized event queue are processed.

\begin{proposition}\label{updates1}
 Let $\strategy \in \qhs$. Let $\strategy' \in \qhs$ be a strategy
 which assigns the highest priority to any updated event which is
 redundant.  Then, $\forall$
 $P \in {\it Prog}$, $\dom \in \adom$, $\atom \subseteq \aatom$,
 \mbox{\analyzerA($\atom,\strategy$)=\analyzerA($\atom,\strategy'$)}.
\end{proposition}

\subsection{The Notion of Reduced Certificate}\label{sec:noti-reduc-cert}

As mentioned above, the notion of reduced certificate is directly
related to the idea of recomputation in the program analysis graph.
Now, we are interested in finding those entries $A:{\it CP} \in \entries$ in the
answer table, whose analysis has launched the reprocessing of some arcs
and hence recomputation has occurred. Certainly, the reprocessing of
an arc may only be caused by a non-redundant updated event for
$A:{\it CP}$, which inserted (via {\sc add\_dependent\_rules}) all
arcs in ${\it DAT|_{A:{\it CP}}}$ into the prioritized event queue. However
some updated events are not dangerous. For instance, if the processing
of an arc $H:{\it CP_0}\Rightarrow[{\it CP_1}]A:{\it CP}$ has been
stopped because of the lack of answer for $A:{\it CP}$ (L\ref{24}
and L\ref{last} in Algorithm \ref{algo:gen-anal}), this arc must be
considered as ``suspended'', since its continuation has not been
introduced in the queue. In particular, we do not take into account
updated events for $A:{\it CP}$ which are generated when ${\it
  DAT}|_{A{:}{\it CP}}$ only contains suspended arcs. Note that this
case still corresponds to the first traversal of any arc and should
not be considered as a reprocessing.  The following definition
introduces the notion of {\em suspended arc}, i.e., of an arc
suspended during analysis.

\begin{definition}[suspended arc] \label{def:suspended-arc}  
  Let $P \in \prog$, $\atom \subseteq \aatom$ and $\Omega \in \qhs$. 
  We say that an arc
  $H:{\it CP_0} \Rightarrow [{\it CP_1}]\ B:{\it CP_2}$ in the
  dependency arc table is {\em suspended} w.r.t. $\Omega$ during the
  analysis of $P$ for $\atom$ iff when it is generated, the answer
  table does not contain any entry for $B:{\it CP_2}$ or contains an
  entry of the form $B:{\it CP_2} \mapsto \bot$.
\end{definition}

For the rest of the updated events, their relevance depends strongly
on the strategy used to handle the prioritized event queue.
For instance, assume that the prioritized event queue contains an event 
${\it arc}(H:{\it CP_0}\Rightarrow [{\it CP_1}]A:{\it CP})$, coming from 
a suspended arc in {\it DAT}. If all updated events for $A :{\it CP}$
are processed before this arc (i.e., the fixpoint of $A :{\it CP}$ is
available before processing the arc), 
then these updated events do not launch any recomputation. Let us define now
the notion of recomputation.

\begin{definition}[multi-traversed arc] \label{def:initial-update}
 Let $P \in \prog$, $\atom \subseteq \aatom$ and $\Omega \in \qhs$. We say that an 
arc $H:{\it CP}\Rightarrow [{\it CP_0}]A:{\it CP}_1$ in the
dependency arc table has been 
{\em multi-traversed} w.r.t. $\Omega$ after the analysis of $P$ for $\atom$ 
 iff it has been introduced in the dependency arc
table at least twice as a non suspended arc w.r.t. $\Omega$.
\end{definition}

\begin{example}\label{ex:4}
  Assume that we use a strategy $\strategy'' \in \qhs$ such that step
  C in Example~\ref{ex:rectoy1} is performed before B (i.e., the second
  rule is analyzed before the first one).  Then, when the answer for
  $\tt rectoy(N1,R){:}\tuple{int,term}$ is looked up, procedure {\sc
    get\_answer} returns $\bot$ and thus the processing of arc
  $A_{2,2}$ is \emph{suspended} at this point in the sense that its
  continuation $A_{2,3}$ is not inserted in the queue (see L\ref{24}
  in Algorithm~\ref{algo:gen-anal}).
  Indeed, we can proceed with the
  remaining arc $A_{1,1}$ which is processed exactly as in step B.  In
  this case, the updated event $U_1$ is not redundant for
  $\strategy''$, as there is a suspended dependency introduced by the former
  processing of arc $A_{2,2}$ in the table.  Therefore, the processing
  of $U_1$ introduces the suspended arc $A_{2,2}$ again in the queue, and
  again $A_{2,2}$ is introduced in the dependency arc table, 
  but now as not suspended.
  The important point is that the fact that $U_1$ inserts $A_{2,2}$
  must not be considered as a reprocessing, since $A_{2,2}$ had been
  suspended and its continuation ($A_{2,3}$ in this case) had not been
  handled by the algorithm yet. Hence, finally  $A_{2,2}$ has not
been multi-traversed. \hfill $\Box$
\end{example}

We define now the notion of {\em relevant entry}, which will
be crucial for defining reduced certificates.
The key observation is that those answer patterns whose computation has
generated multi-traversed arcs should be available in the certificate.

\begin{definition}[relevant entry]  \label{def:relevant-entry}
   Let $P \in \prog$, $\atom \subseteq \aatom$ and $\Omega \in \qhs$. We say that the entry
  $A : {\it CP} \mapsto {\it AP}$ in the answer table is \emph{relevant} 
  w.r.t. $\Omega$ after the analysis of $P$ for $\atom$  iff there
  exists a multi-traversed arc  $\_\ \Rightarrow [\_]A:{\it CP}$
  w.r.t. $\Omega$ in the dependency arc table.
\end{definition}

\noindent
The notion of \emph{reduced certificate} allows us to remove
irrelevant entries from the answer table and produce a smaller
certificate which can still be validated in one pass.

\begin{definition}[reduced certificate]
  \label{def:reduced-certificate}  Let $P \in \prog$, $\atom \subseteq
  \aatom$ and $\Omega \in \qhs$.  Let \certificateA~=
  \analyzerA($\atom,\Omega$) for $P$ and $\atom$.  We define the
  \emph{reduced certificate}, $\certificateB$, as the set of relevant
  entries in $\certificateA$ w.r.t. $\Omega$.
\end{definition}

\begin{example} From now on, in our running example, we assume the
  strategy $\strategy' \in \qhs$ which assigns the highest priority to
  redundant updates (see Proposition \ref{updates1}). For this
  strategy, the entry $E'_1 \equiv \tt \tt
  rectoy(N,M):\tuple{int,term} \mapsto \tuple{int,int}$ in
  Example~\ref{ex:rectoy1} is not relevant since no arc has been
  multi-traversed.
  Therefore, the reduced certificate for our running example is empty.
  In the following section, we show that our checker is able to reconstruct
  the fixpoint in a single pass from the empty certificate. It should
  be noted that, using $\strategy$ as in Example~\ref{ex:rectoy1}, the
  answer is obtained by performing two analysis iterations over the
  arc associated to the second rule of $\tt rectoy(N,M)$ (steps C and
  D) due to the fact that $U_1$ has been delayed and becomes relevant
  for $\strategy$. Thus, this arc has been multi-traversed. \hfill $\Box$

\end{example}

\noindent
Consider now the Java version of the procedure {\tt rectoy},  
borrowed from  \cite{RR98}:

\[ \begin{array}{llll}
\mbox{\sf int {\tt rectoy}(int n) \{} \\
\quad \mbox{\sf int m; int r;} \\
\quad \mbox{\sf m=0;} \\
\quad \mbox{\sf if (n $>$ 0) \{} \\
\quad \quad \mbox{\sf n= n-1;} \\
\quad \quad \mbox{\sf r = this.rectoy(n);} \\
\quad \quad \mbox{\sf m = n + 4;} \\
\quad \mbox{\}};\\
\quad \fbox{\mbox{\sf return m;}}~~~~  \mbox{\tt // Program point 30}\\
\mbox{\sf \}}
\end{array} \]

\noindent
For this program, lightweight bytecode verification (LBV) 
\cite{RR98} sends, together with the
program, the reduced \emph{non-empty} certificate ${\it cert}=( \{30 \mapsto
(\epsilon,rectoy\cdot int \cdot int \cdot  \bot
)\},\epsilon)$, which states that at program point $30$ the stack does
not contain information (first occurrence of $\epsilon$),\footnote{The
  second occurrence of $\epsilon$ indicates that there are no
  backward jumps.} and variables $\sf n$, $\sf m$ and $\sf r$ have
type $\sf int$,
$\sf int$ and $\bot$. The need for sending this information is because
{\tt rectoy}, implemented in Java, contains an \emph{if}-branch
(equivalent to the branching for selecting one of our two clauses for
${\tt rectoy}$). In LBV, $\it cert$ has to inform the checker that it is
possible for variable $\sf r$ at point $30$ to be undefined, if the
\emph{if} condition does not hold. However, in our method this is not
necessary because the checker is able to reproduce this information
itself. Therefore, the above example shows that our approach improves
on state-of-the-art PCC techniques by reducing the certificate even
further while still keeping the checking process one-pass.

%% file: updates.tex

\subsection{Generation of Certificates without Irrelevant Entries}\label{sec:gener-cert-with}

In this section, we  instrument the analyzer of
Algorithm~\ref{algo:gen-anal} with the extensions necessary for
producing reduced certificates, as defined in
Definition~\ref{def:reduced-certificate}.  Together with the answer
table 
returned by Algorithm \ref{algo:gen-anal}, this new algorithm returns also the set 
$\reduced$ (initially empty) of call patterns 
which will form finally the reduced certificate \certificateB. 
 The resulting analyzer
{\analyzerB} is presented in Algorithm \ref{algo:red-anal}. Except for
procedure {\sc process\_arc} and {\sc insert\_answer\_info},
it uses
the same procedures as Algorithm \ref{algo:gen-anal}, adapting them
to the new syntax of arcs. Now, arcs will be annotated with an
integer value $u$ which counts the number of times that the arc
has been traversed during the analysis. The first time that
an arc is introduced in the prioritized event queue, it is annotated with $0$. 
Thus, L\ref{12} in Algorithm \ref{algo:gen-anal} must be replaced 
by:

\[\fbox{\mbox{\ref{12}: \ \ \ 
  {\sf add\_event}($arc(A_k(0):\CP{} \Rightarrow{} [\CP{}_0]
~B_{k,1}:\CP{}_1$),$\strategy$)}} \]

\noindent
Let us see the differences between Algorithm \ref{algo:red-anal}
and Algorithm \ref{algo:gen-anal}:

\begin{enumerate}

\item \emph{We detect all multi-traversed arcs.} When a call to {\sc
    process\_arc} is generated, this procedure checks if the arc is
  suspended (L\ref{suspended3}) before introducing the corresponding
  arc in the dependency arc table.  If the arc is suspended, then its
  $u$ value is not modified, since, as explained before, it cannot be
  considered as a reprocessing. Otherwise, the $u$-value is
  incremented by one. Furthermore, if $B_{k,i}$ is not a constraint
  and $u$ is greater than $1$, then $B_{k,i}{:}{\it CP_2}$ is added to
  the $\reduced$ set, since this means that the arc has been
  multi-traversed.  Note that the $\reduced$ set will contain in the
  end those call patterns whose analysis launches the recomputation of
  some arc.

  Another important issue is how to handle the continuation of the
  arc which is being currently processed. If the arc is suspended, then no
  continuation is introduced in the queue (checked by
  L\ref{suspended1} and L\ref{suspended2}). Otherwise
  (L\ref{suspended1}), before introducing the continuation in the
  queue, we check if the dependency arc table already contains such a
  continuation (L\ref{arc1}). In that case, we add the arc with the
  same $u$ annotation than that in the queue (L\ref{arc33}). Otherwise,
  we introduce the continuation as an arc initialized with $0$
  (L\ref{arc3}).

\item \emph{We ignore redundant updates.} Only non-redundant updates
  are processed by procedure {\sc insert\_answer\_info}
  (L\ref{analucero}).  Each time an updated event is generated, we
  check if ${\it DAT}|_{H{:}CP}$ is different from $\emptyset$
  (L\ref{analucero}).  Only then, an updated event for
  $H{:}{\it CP}$ is generated (L\ref{updatedr}).

\end{enumerate}

\bigskip

\begin{algorithm}[h]
\caption{{\analyzerB}: Analyzer instrumented for Certificate Reduction}

\label{algo:red-anal}\begin{algorithmic} [1]
\small
\Procedure{process\_arc}{$H_k(u):\CP{}_0 \Rightarrow  [\CP{}_1]
  ~B_{k,i}:\CP{}_2 \in \depend$,$\strategy \in \qhs$}
\State $W$ := $vars(H_k,B_{k,1}, \ldots, B_{k,n_k})$; \label{Wcheck1}
\State $\CP{}_3$ := {\sc get\_answer}($B_{k,i}:\CP{}_2, \CP{}_1,
W,\strategy$); \label{get-answer} 
\If {$\CP{}_3 \neq \bot$ and $i \neq n_k$} \label{suspended1}
\State $\CP{}_4$ := {\sf Arestrict}$(\CP{}_3, vars(B_{k,i+1}))$;
\If {there exists the arc $H_k(w):\_ \Rightarrow \_:B_{k,i+1}$ in \newline the
dependency arc table} \label{arc1}
\State ~~{\sf add\_event}($arc(H_k(w):\CP{}_0 \Rightarrow [\CP{}_3]
~B_{k,i+1}:\CP{}_4$),$\strategy$); \label{arc33}
\Else ~{\sf add\_event}($arc(H_k(0):\CP{}_0 \Rightarrow [\CP{}_3]
~B_{k,i+1}:\CP{}_4$),$\strategy$);  \label{arc3}
\EndIf
\ElsIf {$\CP{}_3 \neq \bot$ and $i = n_k$} \label{suspended2}
 \State $\AP{}_1$ := {\sf Arestrict}$(\CP{}_3, vars(H_k))$;
\State {\sc insert\_answer\_info}($H:\CP{}_0 \mapsto
\AP{}_1,\strategy$); \label{end-suspended2}
\EndIf

\If{$B_{k,i}$ is not a constraint} \label{new-forget-check}
\If {${\it CP_3}=\bot$} \label{suspended3}
   \State add $H_k(u){:}\CP{}_0{\Rightarrow }[\CP{}_1]
   ~B_{k,i}{:}\CP{}_2)$ to dependency arc table; \label{20n}
\Else \label{reducedelse}{\tiny~~~~~~~~~\% non-suspended arc}
\State add $H_k(u+1){:}\CP{}_0{\Rightarrow} [\CP{}_1] ~B_{k,i}{:}\CP{}_2$ to dependency arc table; \label{nonsuspendedarc}
\If{$u{+}1{>}1$}~add $B_{k,i}{:}\CP{}_2$ to $\reduced$; \label{arc4}
\EndIf
\EndIf
\EndIf
\EndProcedure
\Statex

\Procedure{insert\_answer\_info}{$H:\CP{} \mapsto \AP{} \in
  \entries,\strategy \in \qhs$}
\State $\AP{}_0$ := {\sc lookup\_answer}$(H:\CP{},\strategy)$;
\State  $\AP{}_1$ := {\sf Alub}$(\AP{},\AP{}_0)$;
\If {$AP_0 \not = AP_1$} \label{analdistintos}  {\tiny~~~~~~~~~\%updated required }
    
\State {\sf add\_answer\_table}($H{:}CP{\mapsto}AP_1$); \label{nosale}

\If {${\it DAT|_{H:CP}}\not =\emptyset$} \label{analucero}{\tiny~~~~~~~~~\% non-redundant updated}
       \State {\sf add\_event}({\it updated}($H:CP$)); \label{updatedr}
   
\EndIf
 
\EndIf
\EndProcedure
\renewcommand{\baselinestretch}{0.4}\scriptsize
\end{algorithmic}
\end{algorithm}

\medskip

\begin{example}\label{ex:anal2}
  Consider the four steps performed in the analysis of our running
  example. Step A is identical. In step B the {\sc
    insert\_answer\_info} procedure detects a redundant updated event
  (L\ref{analucero}). No updated event is generated.  Step C remains
  identical and the arc $A_{2,2}$ (the only one able to contribute to
  the $\reduced$ set) is annotated with $1$, and step D does not
  occur. As expected, upon return, the $\reduced$ set remains
  empty. \hfill $\Box$
\end{example}

\subsection{Correctness of Certification}

This section shows the correctness of the certification process
carried out to generate reduced certificates, based on the
correctness of the certification with full certificates
of~\cite{ai-safety-ngc07}. First, 
note that, except for the control of relevant entries,
\analyzerA($\atom,\strategy$)~and \analyzerB($\atom,\strategy$)~have
the same behavior and thus compute the same answer table.

\begin{proposition}
\label{propestrategia}
Let $P \in \prog$, $\dom \in \adom$, $\atom \subseteq \aatom$, $\strategy ,
\strategy' \in \qhs$.  Let ${\it AT}$ be the answer table computed by
\analyzerB($\atom,\strategy'$).  Then,
\analyzerA($\atom,\strategy$) = ${\it AT}$.
\end{proposition}

\begin{proof} 
First note that except for the $u$-annotations, the procedures
{\sc process\_arc} in Algorithms
\ref{algo:gen-anal} and \ref{algo:red-anal} are similar. In fact, there is a
one-to-one correspondence between the definition of both procedures. 
Concretely, we have the following mapping:

\bigskip

\begin{tabular}{|l | l|}
\mbox{\analyzerA} & \mbox{\analyzerB} \\ 
\mbox{L\ref{20}-L\ref{21}} & \mbox{L\ref{new-forget-check}-\ref{arc4}} \\
\mbox{L\ref{Wcheck}} & \mbox{L\ref{Wcheck1}} \\ 
\mbox{L\ref{23}} & \mbox{L\ref{get-answer}} \\
\mbox{L\ref{notlast}-L\ref{add-event-1}} & \mbox{L\ref{suspended1}-L\ref{arc3}}\\
\mbox{L\ref{last}-L\ref{relast}} & \mbox{L\ref{suspended2}-L\ref{end-suspended2}}\\
\end{tabular} 

\bigskip
\noindent
The only difference between  Algorithms
\ref{algo:gen-anal} and \ref{algo:red-anal} relies on  
{\sc insert\_answer\_info}. For the case of Algorithm
\ref{algo:red-anal}, redundant updates are never introduced in the
prioritized event queue (L\ref{analucero}). Then, let us choose a new strategy
$\Omega''$, identical to $\Omega'$ except when dealing with redundant
updates. For redundant updates, let us assume that $\Omega''$ processes
them inmediately after being introduced in the event
queue. Such processing does not generate any effect since the
dependency arc table does not contain arcs to be launched for these
updates. 
Hence it holds
that \analyzerB($\atom,\Omega'$) generates the same answer table $\it
AT$ than
\analyzerA($\atom,\Omega''$).
From Proposition~\ref{updates1} it holds 
that \analyzerA($\atom,\strategy$)=\analyzerA($\atom,\strategy''$)
and the claim follows.  
\end{proof}

\noindent
The following definition presents the certifier for reduced certificates.

\begin{definition} \label{certreduced}
  We define the function \certifierB: $\prog {\times} \adom {\times} 2^{\aatom}
  {\times} \aint {\times} \qhs {\mapsto} \acert$, which takes $P \in \prog$,
  $\dom \in \adom$, $\atom \subseteq \aatom$,  $\inten
  \in \aint$, $\strategy \in \qhs$. It returns as certificate, 
  $\certificateB = \{A : {\it CP} \mapsto {\it AP} \in \certificateA \ | 
\ A : {\it CP} \in \reduced\}$, 
   where
  $\langle \certificateA,\reduced \rangle = $\analyzerB($\atom,\strategy$), iff $\certificateA
  \sqsubseteq \inten$.
\end{definition}
Finally, we can establish the correctness of \certifierB~which amounts
to say that $\certificateB$ contains all 
relevant entries in $\certificateA$.

\begin{theorem} \label{propU}
Let $P \in Prog$, $\dom \in \adom$, $\atom \subseteq \aatom$, $\inten
\in \aint$ and $\strategy   \in \qhs$. Let
 $\certificateA=$\analyzerA($\atom,\strategy$) and  
 $\certificateB$= \certifierB($P,\dom,\atom,\inten,\strategy$).  
 Then, an entry   $A:{\it CP}\mapsto{\it AP} \in \certificateA$ is relevant w.r.t.  $\strategy$ iff 
  $A:{\it CP}   \mapsto{\it AP} \in \certificateB$.
\end{theorem}

\begin{proof}

According to Definition \ref{certreduced},
$\certificateB = \{A : {\it CP} \mapsto {\it AP} \in \certificateA \ | \ A : {\it CP} \in \reduced\}$, 
   where
  $\langle \certificateA,\reduced \rangle =
  $\analyzerB($\atom,\strategy$). Hence, it is enough to
 prove that an entry
  $A:{\it CP}\mapsto{\it AP} \in \certificateA$ is relevant w.r.t.  $\strategy$ iff 
  $A:{\it CP} \in \reduced$.

\begin{description}
\item[$(\Leftarrow)$] Assume that $A : {\it CP} \in \reduced$. Then,
from L\ref{nonsuspendedarc} and L\ref{arc4} 
 it holds that there exists an arc
$H(u) : {\it CP'}{\Rightarrow}[\_]A : {\it CP}$ in the dependency arc
table such that $u > 1$. But the $u$-value of an arc can only be
increased in procedure {\sc process\_arc} (L\ref{nonsuspendedarc})
after checking that ${\it CP_3}$ is different from $\bot$
(L\ref{reducedelse}). But ${\it CP_3}$ is computed by means of {\sc
  get\_answer} (L\ref{get-answer}) which calls  {\sc
  lookup\_answer}  (L\ref{32}). This last function only returns a value different
from $\bot$ if $A:{\it CP}$ as an entry in the answer table (L\ref{36}
and L\ref{36-1}). Since $u>1$ then $u$ has been incremented at least
twice and as argued before, in both cases the answer table contained
an entry for ${\it A:CP}$, i.e., by Definition
\ref{def:initial-update}, the arc 
$H : {\it CP'} \Rightarrow [\_]A : {\it CP}$ is multi-traversed w.r.t $\Omega$.
Hence, by Definition \ref{def:relevant-entry},
$A : {\it CP} \mapsto {\it AP}$ is a relevant entry.

\item[$(\Rightarrow)$] Assume now that the entry $A :{\it
    CP}\mapsto {\it AP}$ is relevant w.r.t $\Omega$. Then, by Definition
  \ref{def:relevant-entry}, there exists an arc $H:{\it
    CP'} \Rightarrow [\_]A:{\it CP}$ in the dependency arc table which
  has been multi-traversed. By Definition \ref{def:initial-update},
  this arc has been introduced in the dependency arc table at least
  twice as non-suspended arc. But arcs are introduced in ${\it DAT}$
  via procedure {\sc process\_arc} and each time the arc is non
  suspended (L\ref{reducedelse}) its $u$-value is increased by $1$
  (L\ref{nonsuspendedarc}). Hence the $u$ value for
  $H:{\it CP'} \Rightarrow [\_]A : {\it CP}$ is at least $2$. Now,
  L\ref{arc4} ensures that $A:{\it CP} \in \reduced$.
\end{description}
\end{proof}

%% file: checking.tex

\section{Checking Reduced Certificates}\label{sec:generic-checker-full}

In the ACC framework for full
certificates~\cite{lpar04-ai-safety-long} a concrete checking
algorithm is used with a specific graph traversal strategy which we
will refer to as $\strategy_{C}$. This checker has been shown to be
very efficient (i.e., this particular $\strategy_{C}$ is a good
choice) but here we would like to consider a more generic design for
the checker in which it is parametric on $\strategy_{C}$ in addition
to being parametric on the abstract domain.\footnote{Note that both
  the analysis and checking algorithms are always parametric on the
  abstract domain. This genericity allows proving a wide variety of
  properties by using the large set of available abstract domains,
  this being one of the fundamental advantages of ACC.}  This lack of
parametricity on $\strategy_{C}$ was not an issue in the original
formulation of ACC in~\cite{lpar04-ai-safety-long} since there full
certificates were used. Note that even if the certifier uses a
strategy $\strategy_{A}$ which is different from $\strategy_{C}$, all
valid full certificates are guaranteed to be validated in one pass by
that specific checker, independently of $\strategy_{C}$. This result
allowed using a particular strategy in the checker without loss of
generality.  However, the same result does not hold any more in the
case of reduced certificates.  In particular, \emph{completeness} of
checking is not guaranteed if $\strategy_{A} \neq \strategy_{C}$. This
occurs because, though the answer table is identical for all
strategies, the subset of redundant entries depends on the particular
strategy used. The problem is that, if there is an entry $A : CP
\mapsto {\it AP}$ in $\certificateA$ such that it is relevant w.r.t.\
$\strategy_{C}$ but it is not w.r.t.\ $\strategy_{A}$, then a
single-pass checker will fail to validate the $\certificateB$
generated using $\strategy_{A}$. In this section, we design a generic
checker which is not tied to a particular graph traversal strategy.
In practice, upon agreeing on the appropriate parameters, the consumer
uses the particular instance of the generic checker resulting from the
application of such parameters.  In a particular application of our
framework, we expect that the graph traversal strategy is agreed a
priori between consumer and producer.  Alternatively, if necessary
(e.g., when the consumer does not implement this strategy), the
strategy can be sent along with the certificate in the transmitted
package.

It should be noted that the design of generic checkers is also
relevant in light of current trends in verified analyzers (e.g.,
\cite{verified-KN03,CJPR04-long}), which could be transferred directly to
the checking end.  In particular, since the design of the checking
process is generic, it becomes feasible in ACC to use automatic
program transformation techniques \cite{pevalbook93} to specialize a
certified (specific) analysis algorithm in order to obtain a certified
checker with the same strategy while preserving correctness and
completeness.

\subsection{The Generic Checking Algorithm}

The following definition presents a generic checker for validating
reduced certificates.
In addition to the genericity issue discussed above, an important
difference with the checker for full certificates
\cite{lpar04-ai-safety-long} is that there are certain entries which are
not available in the certificate and that we want to reconstruct and
output in checking. The reason for this is that the safety policy has
to be tested w.r.t.\ the full answer table --Equation~(\ref{eq:2}).
Therefore, the checker must reconstruct, from \certificateB, the
answer table returned by \analyzerA, \certificateA, in order to test
for adherence to the safety policy --Equation~(\ref{eq:4}).  Note that
reconstructing the answer table does not add any additional cost
compared to the checker in \cite{lpar04-ai-safety-long}, since the full
answer table also has to be created in~\cite{lpar04-ai-safety-long}.

\bigskip

\begin{algorithm}[h]
      \caption{Generic Checker for Reduced Certificates \checkingB}
\label{algo:red-check}\begin{algorithmic} [1]
\small
\Procedure{insert\_answer\_info}{$H{:}\CP{}{\mapsto}\AP{} \in
  \entries,\strategy \in \qhs$}
\State $\AP{}_0$ := {\sc lookup\_answer}$(H{:}\CP{},\strategy)$;
\State $\AP{}_1$ := {\sf Alub}$(\AP{},\AP{}_0)$;\label{lub1}
\State (${\it IsIn}$,$AP'$)={\sc look\_fixpoint}($H{:}CP$,\certificateB); \label{lookfix}
 \If{${\it IsIn}$ and {\sf Alub}$(\AP{},\AP{}') \not = AP'$}~return~\error;
 {\footnotesize~~~~~~~~~\% error of type a) }
 \EndIf \label{checkinitialerror} \label{initialisIn} \label{checkrelevantin}

\If {$AP_0{\not=}AP_1$} \label{checkinitial}{\footnotesize~~~~~~~~~\%
  updated required}
\If{${\it IsIn}$ and $AP_0{=}\bot$} $AP_1{=}AP'$ \EndIf \label{putfixpoint}
\State {\sf add\_answer\_table}($H{:}{\it CP}{\mapsto}{\it AP_1}$);   \label{analguardaanswerc} \label{checkanswer1} \label{checknotisin}
  \If {${\it DAT|_{H:CP}}\not{=}\emptyset$}
   \State {\sf  add\_event}(${\it updated}(H{:}CP),\strategy$);  \label{checkupdate} \label{checkucero}
    \EndIf
       
 \label{relevanterror}
\EndIf
\EndProcedure

\Statex
\Function{look\_fixpoint}{$A{:}{\it CP} \in \aatom$, \certificateB~ $\in
  \acert$}
\If{$\exists$
  a renaming $\sigma$ such that $\sigma(A{:}{\it CP}{\mapsto}{\it AP}){\in}$
  \certificateB} 
  \State return (\true,$\sigma^{-1}({\it AP})$);
 \Else~return (\false,$\bot$);
\EndIf
\EndFunction
\Statex

         \renewcommand{\baselinestretch}{0.4}\scriptsize

       \end{algorithmic}

 \end{algorithm}

\begin{definition}[checker for reduced certificates]
\label{def:reduced-checker}

  Function \textsc{\checkingB} is defined as function \analyzerB~with
  the following modifications:

\begin{enumerate}

\item It receives \certificateB~as an additional input parameter.

\item It does not use the set $\reduced$ and it replaces L\ref{arc4} of
  Algorithm \ref{algo:red-anal} with: 

\[\fbox{\mbox{\ref{arc4}: 
  If  $u{+}1{>}1$~{\bf
  return}~\error}} 
\]

\item If it fails to produce an answer table, then it issues an
  \error.

  \item Function {\sc insert\_answer\_info} is replaced by the new
    one in Algorithm~\ref{algo:red-check}.

  \end{enumerate}

\noindent
Function \textsc{\checkerB}~takes $P \in Prog$, $\dom \in \adom$,
$\atom \subseteq \aatom$, $\inten \in \aint$, $\strategy \in \qhs$,
$\certificateB \in \acert$ and returns:

\begin{enumerate}

\item \error~if \checkingB($\atom,\strategy,\certificateB$) for $P$ in
  $\dom$ returns \error.

\item Otherwise it returns $\certificateA$=\checkingB($\atom,\strategy,\certificateB$) for $P$ and $\dom$ iff
$\certificateA \sqsubseteq \inten$.
\end{enumerate}

\end{definition}

\noindent
Let us briefly explain the differences between Algorithms
\ref{algo:red-anal} and \ref{algo:red-check}.  First, the checker has
to detect (and issue) two sources of errors:

\begin{itemize}
\item[a)] The answer in the certificate and the one
  obtained by the checker differ (L\ref{initialisIn}).
  This is the traditional error in ACC and means that the certificate
  and program at hand do not correspond to each other. The
call to function {\sc  look\_fixpoint}($H:{\it CP}$,\certificateB) in
L\ref{lookfix}
returns
a tuple (${\it
  IsIn},AP'$) such that: if $H : {\it CP}$ is in \certificateB, then ${\it
  IsIn}$ is equal to \true~and $AP'$ returns the fixpoint stored in
\certificateB. Otherwise, ${\it IsIn}$ is equal to \false~and $AP'$ is
$\bot$.

\item[b)] Recomputation is required.  This should not occur during
  checking, i.e., no arcs must be multi-traversed by the checker
  (L\ref{arc4}). This second type of error corresponds to situations
  in which some non-redundant update is needed in order to obtain an
  answer (it cannot be obtained in one pass).  This is detected in
  L\ref{arc4} prior to check that the arc is not suspended
  (L\ref{suspended2}) and it has been traversed before, i.e., its $u$
  value is greater than 1.  Note that we flag this as an error because
  the checker will have to iterate and the description we provided
  does not include support for it. In general, however, it is also
  possible to use a checker that is capable of iterating. In that case of
  course the certificates transmitted can be even smaller than the
  reduced ones, at the cost of increased checking time (as well as
  some additional complexity in the checking code).  This allows
  supporting different tradeoffs between certificate size, checking
  time, and checker code complexity.

\end{itemize}

\noindent The second difference is that the $A : {\it
  CP} \mapsto {\it AP'}$ entries 
stored in \certificateB\ have to be added to the answer table after finding
the first partial answer for $A : {\it CP}$ (different from $\bot$),
in order to detect errors
of type a) above.  In particular, L\ref{putfixpoint} and
L\ref{analguardaanswerc} add the fixpoint ${\it AP}'$ stored in
\certificateB~to the answer table.

\begin{example}
All steps given for the analysis of Example~\ref{ex:anal2} are identical in
 \checkerB\, except for the detection of possible errors. Errors
of type a) are not possible since \certificateB~is
 empty. An error of type b) can only be generated because of the
$u$ value of arc $A_{2,2}$. However note that in step C, this arc is
introduced in the queue with $u=0$. After processing the arc, the arc
goes to the dependency arc table with $u=1$. But since no updated
events are generated, this arc is no longer processed.  Hence, 
 the program is validated in a single pass over the
 graph. \hfill $\Box$

\end{example}

\subsection{Correctness of Checking}

In this section we prove the correctness of the checking process,
which amounts to saying that if \checkerB\ does not issue an error
when validating a certificate, then the reconstructed answer table
is a fixpoint verifying the given input safety policy. As a previous
step, we prove the following proposition in which we also ensure that
the validation of the certificate is done in one pass.

 \begin{proposition} \label{corr}
 Let $P \in {\it Prog}$, $\dom \in \adom$, $\atom \subseteq \aatom$,
   $\inten \in \aint$ and $\strategy \in \qhs$.  Let
   \certificateA =   \certifierA($P,\dom,\atom,\inten,\strategy$),
   \certificateB=\certifierB($P,\dom,\atom,\inten,\strategy$). Then
   \checkingB($\atom,\strategy,\certificateB$) does not
   issue an \error~and it returns \certificateA. Furthermore, the validation
 of \certificateA~does not generate multi-traversed arcs.

 \end{proposition}

\begin{proof}
Let us consider first the call:

\[(*)~\mbox{\checkingB}(\atom,\strategy,\certificateB) \]

\noindent
For this call, let us prove that (1) it does not issue an error and;
(2) it returns $\certificateA$ as result.

\bigskip

\noindent
{\bf (1)
  \mbox{\checkingB}($\atom,\strategy,\certificateB$) 
   does not issue an error.}

\bigskip

\noindent
Errors of type (a) (L\ref{initialisIn} of Algorithm
 \ref{algo:red-check}) are not possible since, from Definition
 \ref{certreduced}, $\certificateB \subseteq
 \certificateA$, where $\certificateA$ is the answer table computed by
 \analyzerA($\atom,\strategy$). The correctness of 
  Algorithm \analyzerA($\atom,\strategy$) (see \cite{incanal-toplas}) avoids this
  kind of errors.

  Errors of type (b) can only occur in L\ref{arc4} of procedure {\sc
    process\_arc} (Algorithm \ref{algo:red-check}), for some arc
  $H:{\it CP_0} \Rightarrow [{\it CP_1}] B_{k,i}:{\it CP_2}$. Since we
  follow the same strategy $\Omega$ in \checkingB~and \analyzerB,
  then \analyzerB($\atom,\Omega$) introduces $B_{k,i}:{\it CP_2}$ in
  {\sf RED} (L\ref{arc4} of Algorithm \ref{algo:red-anal}), and thus,
  Definition \ref{certreduced} ensures that $B_{k,i}:{\it CP_2}
  \mapsto {\it AP} \in \certificateB$. But this is a
  contradiction since for all entries in $\certificateB$,
  the first time that the arc is processed without answer in ${\it
    AT}$ for $B_{k,i}:{\it CP_2}$, Algorithm \ref{algo:red-check} 
  (L\ref{putfixpoint} and L\ref{checkanswer1}) introduces $B_{k,i}:{\it
    CP_2} \mapsto {\it AP}$ in $\it AT$ together with the
  corresponding event ${\it updated(B_{k,i}:{\it CP_2})}$. So when
  $\Omega$ selects this event, the new event ${\it arc}(H:{\it CP_0}
  \Rightarrow [{\it CP_1}] B_{k,i}:{\it CP_2})$ is again introduced in
  the prioritized event queue. When this arc is selected by
  $\Omega$, the arc goes again to ${\it DAT}$. But since
  $B_{k,i}:{\it CP_2} \mapsto {\it AP} \in {\it AT}$, no more events
  of the form $\it updated(B_{k,i}:CP_2)$ may occur
  (L\ref{checkinitial} of {\sc insert\_answer\_info}($B_{k,i}:{\it
    CP_2}$) in Algorithm \ref{algo:red-check} never holds). Hence, no
  more calls to process arc for ${\it arc}(H:{\it CP_0} \Rightarrow
  [{\it CP_1}] B_{k,i}:{\it CP_2})$ occur. Then the $u$-value for this
  arc will be at most $1$ and no error will be generated.

\bigskip

\noindent
{\bf (2) The call $(*)$ returns \certificateA}.

\bigskip

\noindent
The only differences
between the call $(*)$ and the call \analyzerB($\atom,\Omega$) rely on
procedure {\sc insert\_answer\_info} and L\ref{arc4} 
of procedure {\sc process\_arc}.  Since (1) ensures that no error is
issued by $(*)$, then L\ref{checkinitialerror} and L\ref{arc4} of Algorithm
\ref{algo:red-check} are never executed. Then, it is trivial that (1)
computes an answer table $\it AT$ as result. Furthermore, since $(*)$ and 
\analyzerB($\atom,\Omega$) use the same strategy, the only
difference is in the prioritized event queue since for $(*)$ no
relevant updates will appear in the queue. Instead of this, the real
fixpoints in $\certificateB \subseteq \certificateA$ are
introduced in $\it AT$ in L\ref{putfixpoint} and L\ref{checkanswer1} of {\sc
  insert\_answer\_info}. Except for this fact, Algorithms
\ref{algo:red-anal} and \ref{algo:red-check} behave identically and thus
$(*)$ computes $\certificateA$ as result.

\bigskip

Finally, proving that the validation of $\certificateB$
does not generate multi-traversed arcs is trivial since, by
definition, multi-traversed arcs correspond to arcs in ${\it DAT}$
with the $u$-value greater than $1$.  Since the call $(*)$ does not
issue an error, L\ref{arc4} of Algorithm \ref{algo:red-check} is never
executed, i.e., no arc is multi-traversed.

\end{proof}

 \begin{corollary} \label{corr-coro}
 Let $P \in {\it Prog}$, $\dom \in \adom$, $\atom \subseteq \aatom$,
   $\inten \in \aint$ and $\strategy \in \qhs$.  Let
   \certificateA =\certifierA($P,\dom,\atom,\inten,\strategy$),
   and \certificateB$_{\strategy}$=\certifierB($P,\dom,\atom,\inten,\strategy$). If
   \checkingB($\atom,\strategy,\certificateB$), 
   $\certificateB \in \acert$, does not
   issue an \error, then it returns \certificateA~and
   $\certificateB_{\Omega} \subseteq \certificateB$. Furthermore, the validation
 of \certificateA~does not generate multi-traversed arcs.

 \end{corollary}

\begin{proof}
Let us prove, by contradiction, 
that $\certificateB_{\Omega} \subseteq \certificateB$. If
we
assume that $\certificateB_{\Omega} \not \subseteq
\certificateB$, then there exists an entry $A:{\it CP}
\mapsto {\it AP} \in \certificateB_{\Omega}$ such that $A:{\it CP}
\mapsto {\it AP} \not \in \certificateB$. By definition of
$\certificateB_{\Omega}$, $A:{\it CP} \in {\sf RED}$. Hence, L\ref{nonsuspendedarc}
of Algorithm \ref{algo:red-anal} ensures that there exists an arc
$H: {\it CP_0}(u) \Rightarrow [{\it CP_1}] A:{\it CP}$ in ${\it DAT}$
with $u > 1$. But this is not possible since otherwise
the  call \checkingB($\atom,\strategy,\certificateB_{\Omega}$) 
would issue an error, what is a contradiction by Proposition \ref{corr}.

\bigskip

Now observe that from Proposition \ref{corr} it holds that
\checkingB($\atom,\Omega,\certificateB_{\Omega}$) returns
$\certificateA$ and the validation of $\certificateB_{\Omega}$
does not generate multi-traversed arcs. 
But since $\certificateB_{\Omega} \subseteq
\certificateB$, then it trivially holds that 
\checkingB($\atom,\Omega,\certificateB$) also returns $\certificateA$
exactly in the same way that
\checkingB($\atom,\Omega,\certificateB_{\Omega}$) does, i.e., without
generating multi-traversed arcs. 
\end{proof}

\begin{theorem}[correctness] \label{corr}
   Let $P \in {\it Prog}$, $\dom \in \adom$, $\atom \subseteq \aatom$,
   $\inten \in \aint$, $\strategy \in\qhs$ and 
   $\certificateB \in \acert$. Then, if \checkerB($P,\dom,\atom,\inten,\Omega,\certificateB$) does not issue an
   \error~and returns a certificate $\certificateA \in \acert$, then

\begin{itemize}
\item $\certificateA$ is a fixpoint of $P$.
\item $\certificateA \sqsubseteq \inten$;
\end{itemize}
   
\end{theorem}

\begin{proof} If  
   \checkerB($P,\dom,\atom,\inten,\Omega,\certificateB$) does not issue an
   \error~then, from  Definition~\ref{def:reduced-checker}, it holds
   that $\certificateA=$\checkingB($\atom,\Omega,\certificateB$)  
   does not issue an error and $\certificateA \sqsubseteq \inten$. 
   From Corollary \ref{corr-coro}. it follows that
   $\certificateA = $\certifierA($P,\dom,\atom,\inten,\Omega)$.
   Hence, as Definition \ref{def:gen-cert} establishes,  
   $\certificateA$ is the answer table
    computed by \analyzerA($\atom,\Omega'$). Finally, by the results
    in
     \cite{incanal-toplas},  $\certificateA$ is a fixpoint for $P$.

\end{proof}

\subsection{Completeness of Checking}

\noindent The following theorem (completeness) provides sufficient
conditions under which a checker is guaranteed to validate reduced
certificates which are actually valid.  In other words, if a
certificate is valid and such conditions hold, then the checker is
guaranteed to validate the certificate. Note that it is not always the
case when the strategy used to generate it and the one used to check
it are different.

\begin{theorem}[completeness]\label{th:comp}
  Let $P \in \prog$, $\dom \in \adom$, $\atom \subseteq \aatom$, $\inten \in
  \aint$ and $\strategy_A \in \qhs$.  Let \certificateA {=}
  \certifierA($P,\dom,\atom,\inten,\strategy_A$) and
  \certificateB$_{\strategy_A}${=}
  \certifierB($P,\dom,\atom,\inten,\strategy_A$).  Let $\strategy_C
  {\in}$ $\qhs$ be such that  \certificateB$_{\strategy_C}${=}
  \certifierB($P,\dom,$ $\atom,\inten,\strategy_C$) and
  \certificateB$_{\strategy_A}${$\supseteq$}
  \certificateB$_{\strategy_C}$. Then,
  \checkerB($P,$ $\dom,\atom,\strategy_C,\certificateB_{\strategy_A}$)
  returns \certificateA\ and does not issue an \error.

\end{theorem}

\begin{proof}
  We prove it by contradiction. The only cases in which
  \checkerB($P,\dom,\atom,$ $\strategy_C,\certificateB_{\strategy_A}$) issues an \error~are the
  following:

\begin{itemize}
\item The partial answer $\it AP$ computed for some calling pattern $A
  :{\it CP}$
  (provided in \certificateB$_{\strategy_A}$) leads to ${\sf
    Alub}(AP,AP')\not = AP'$ (L\ref{checkinitialerror}), where $AP'$
  is the answer for $A : {\it CP}$, i.e., $A : {\it CP} \mapsto {\it AP'} \in$
  \certificateB$_{\strategy_A}$.  But, 
  \certificateB$_{\strategy_A} \subseteq  \certificateA$, i.e.,
  \certificateA~would contain an incorrect answer for $A : {\it CP}$,
  which is a contradiction with 
  the assumption that \certificateA~is a valid certificate for $P$.

\item There exists some arc $H:{\it CP} \Rightarrow [{\it
    CP_1}]B : {\it CP_2}$ which has been traversed more than once,
  i.e., its $u$-value is greater than $1$ (L\ref{arc4} in Algorithm
  \ref{algo:red-check}). Since $\certificateB_{\strategy_C} \subseteq
  \certificateB_{\strategy_A}$, i.e., $\certificateB_{\strategy_C}$
  contains possibly less entries than $\certificateB_{\strategy_A}$,
  then the call $(*)$ in Theorem \ref{corr} fails also because of such
  a multi-traversed arc. But this is a contradiction with ${\bf (1)}$ in
  Theorem \ref{corr}. 
\end{itemize}

\noindent
Consequently, \checkerB($P,D,S,\strategy_C,\certificateB_{\strategy_A}$) returns an answer table
$\it AT$.  Finally, by Theorem \ref{corr}, we know that since no \error\
is issued, then \checkerB~returns \certificateA.  
\end{proof}

Obviously, if $\strategy_C=\strategy_A$ then the checker is guaranteed
to be complete. Additionally, a checker using a different strategy
$\strategy_C$ is also guaranteed to be complete as long as the
certificate reduced w.r.t $\strategy_C$ is equal to or smaller than
the certificate reduced w.r.t $\strategy_A$. Furthermore, if the
certificate used is full, the checker is complete for any strategy.
Note that if \certificateB$_{\strategy_A}$ $\not \supseteq$
  \certificateB$_{\strategy_C}$, \checkerB\ with the strategy
$\strategy_C$  may fail to validate  \certificateB$_{\strategy_A}$, which is
indeed valid for the program under $\strategy_A$.

\begin{example}
  Consider the program of Example~\ref{ex:informal}, the same abstract
  domain $\dom$ than in our running example, and the call pattern
  $\atom=\{\tt q(X){:}\tuple{term}\}$:
  The full certificate computed by \certifierA~is $\certificateA =
  \{{\tt q(X){:}\tuple{term} \mapsto \tuple{real},}$ ${\tt
    p(X){:}\tuple{term} \mapsto \tuple{real}}\}$.  Let us consider two
  different queue handling strategies $\strategy_A \neq \strategy_C$.
 Under both strategies, we start the analysis introducing
 $\tt q(X){:}\tuple{term} \mapsto \bot$   in the
  answer table and processing the single rule for
  {\tt q}. The arc  
  $\tt q(X)(0){:}\tuple{term}{\Rightarrow}[\{X/term\}] \ p(X){:} \tuple{term}$ is introduced in the queue and processed afterward. As a result,
${\tt q(X)(0)}{:}\tuple{{\tt term}}$ ${\Rightarrow}[{\tt \{X/term\}}]
\ {\tt p(X)}{:} \tuple{\tt term}$ goes to $\it DAT$ and  event ${\it newcall}$({\tt
    p(X){:}$\tuple{\tt term}$}) is generated. The processing of this
  last event adds ${\tt p(X){:}\tuple{term}}$
  ${\tt \mapsto \bot}$ to the answer table.   Now, using
  $\strategy_A$, the analyzer processes both  rules for $\tt p(X)$ in
  textual order. None of the arcs introduced in $\it DAT$ can issue an 
  error.  After traversing the
  first rule, answer $\tt p(X){:}\tuple{term} \mapsto \tuple{real}$
  is inferred and non-redundant updated event 
  ${\it updated}$($\tt p(X){:}\tuple{term}$) is generated.  The
  analysis of the second rule produces as answer $\tt \tuple{int}$ and
  does not update the entry since $\tt \alub(\{X/real\},\{X/int\})$
  returns $\tt \{X/real\}$. We process the non-redundant update for {\tt p} by
  calling function {\sc add\_dependent\_rules}.  The arc for ${\tt q}$
  stored in the dependency arc table with $0$ 
  is launched. When processing this arc, again the arc is introduced
  in $\it DAT$ with $u=1$, and the answer $\tt q(X){:}\tuple{term}
  \mapsto \tuple{real}$ 
   replaces the old one in the answer table. Since $\reduced$ is empty, then 
  \certificateB$_{\strategy_A}$~is empty.

  Assume now that $\strategy_C$ assigns a higher priority to the
  second rule of $\tt p$. In this case, the answer for $\tt
  p(X){:}\tuple{term}$ changes from $\bot$ to $\tt \{X/int\}$, producing
  a non-redundant update.
  Suppose
  now that the updated event is processed, which launches the arc
  for $\tt q$ stored in $\it DAT$. If we process such an arc, then
  it will be introduced again in $\it DAT$, but now with $u=1$. 
  Answer ${\tt q(X){:}\tuple{term} \mapsto
    \{X/int\}}$ is inserted in the answer table.
  When the first arc for ${\tt p}$ is processed, the computed
  answer is $\tt \{X/real\}$. Now, a new non-redundant updated event
 is needed. 
  The processing of this update event launches again the arc for $\tt q$ stored in $\it DAT$, whose analysis introduces it in $\it DAT$ with $u=2$.

  Hence   \certificateB$_{\strategy_A}$ 
    is empty but
    \certificateB$_{\strategy_C}$  
  contains the single entry $\tt
  p(X){:}\tuple{term} \mapsto \tuple{real}$. Thus,
  \checkerB($P,$ $\dom,\atom,$ $\strategy_C,\certificateB_{\strategy_A}$)
  will issue an error (L\ref{arc4}) when trying to validate the
  program if provided with the empty certificate $\certificateB_{\strategy_A}$.
  On the contrary, by Theorem \ref{th:comp}, \checkerB($P,$
  $\dom,\atom,\certificateB_{\strategy_C},\strategy_A$) returns
  \certificateA\ and does not issue an \error.  This
  justifies the results intuitively shown in
  Section~\ref{sec:informal}. \hfill $\Box$
\end{example}

%% file: experiments.tex

\section{Discussion and Experimental
  Evaluation}\label{sec:experiments}

As we have illustrated throughout the paper, the reduction in the size
of the certificates 
is directly related to the number of \emph{updates} (or iterations)
performed during analysis. Clearly, depending on the
``quality'' of the 
graph traversal strategy used, different instances of the generic
analyzer will generate reduced certificates of different sizes.
Significant and successful efforts have been made during recent years
towards improving the efficiency of analysis. The most optimized
analyzers actually aim at reducing the number of updates necessary to
reach the final fixpoint~\cite{inc-fixp-sas-semi-short}. Interestingly, our
framework greatly benefits from all these advances, since the more
efficient analysis, the smaller the corresponding reduced
certificates. We have implemented a generator and a checker of
reduced certificates
as an extension of the efficient, highly optimized, state-of-the-art
analysis system available
in \ciaopp.
Both the analysis and checker use the optimized depth-first
new-calling QHS of~\cite{inc-fixp-sas-semi-short}.

In our experiments 
we study
two crucial points for the practicality of our proposal: the size of
reduced vs.\ full certificates 
(Table~\ref{tab:tableCert}) 
and the
relative efficiency of checking reduced vs.\ full certificates 
(Table~\ref{tab:tableAnCh}).  
As mentioned before, the 
algorithms are parametric w.r.t.\ the abstract domain. In all our
experiments we use the same implementation of the domain-dependent
functions of the {\em sharing+freeness}~\cite{freeness-iclp91}
abstract domain.  We have selected this domain because it is highly
optimized and also because the information it infers is very useful for
reasoning about instantiation errors, which is a crucial aspect for
the safety of logic programs. Furthermore, as mentioned previously, 
sharing domains have also
been shown to be useful for checking properties of imperative
programs, including for example 
information flow characteristics of Java
bytecode~\cite{spoto:pair_sharing,GenaimS05-semi-short}. 
On the other hand, we have used $\top$
as call patterns in order to get all possible modes of use of
predicate calls.

The whole system is written in
\ciao~\cite{ciao-reference-manual-1.13-short} and the experiments have
been run using version 1.13r5499 with
compilation to bytecode on a 
Pentium 4 (Xeon) at 2 Ghz and with 4 Gb of RAM, running GNU Linux Fedora
Core-2 2.6.9.

A relatively wide range of programs has been used as benchmarks. They
are the same ones used in~\cite{incanal-toplas,lpar04-ai-safety-long},
where they are described in more detail. 

\subsection{Size of Reduced Certificates}
\input{tablaN2.tex}

Table~\ref{tab:tableCert} shows our experimental results regarding
certificate size reduction, coded in compact (\emph{fastread}) format,
for the different benchmarks. It compares the size of each reduced
certificate to that of the full
certificate and to the corresponding source code for the same program.

The column \textbf{Source} shows the size of the source code and
\textbf{ByteC} its corresponding bytecode. To make this comparison
fair, in column \textbf{BC/S} we subtract 4180 bytes from
the size of the bytecode for each program: the size of the bytecode
for an empty program in this version of \ciao (minimal top-level
drivers and exception handlers for any executable).  
The size of the certificates is showed in the following  columns. The
columns \textbf{FCert} and \textbf{RCert} contain the size of the full
and reduced certificates, respectively, for each benchmark, and they are
compared in the next column (\textbf{F/R}). Our results show that the
reduction in size is quite significant in all cases.  It ranges from
$101.45$ in ${\it zebra}$ (\certificateB~is indeed empty --the minimum
size of an empty certificate is 40 bytes-- whereas \certificateA~is
$4058$) to $1$ for ${\it deriv}$ (both certificates have the same
size).

The last column (\textbf{R/S}) compares the size of the reduced
certificate to the source code (i.e., the size of the final package
to be submitted to the consumer).  The results show the size of the
reduced certificate to be very reasonable.  It ranges from 0.018 times
the size of the source code (for zebra) to 1.144 (in the case of
serialize).  Overall, it is 0.28 times the size of the source code.  We
consider this satisfactory since in general (C)LP programs are quite
compact (up to 10 times more compact than equivalent imperative
programs). 

\subsection{Checking Time of Reduced Certificates}

\input{tablaN1alt.tex}

Table~\ref{tab:tableAnCh} presents our experimental results regarding
checking time.  Execution times are given in milliseconds and measure
\emph{runtime}.  They are computed as the arithmetic mean of five
runs. 
For each benchmark, 
columns
\textbf{C$_F$} and \textbf{C$_R$} are the times for executing
\checkerA~and \checkerB, respectively. Column \textbf{C$_F$/C$_R$}
compares both checking times.
These times show that the efficiency of \checkerB\ is very similar to
that of \checkerA\ in most cases.

The last row (Overall) summarizes the results for the different
benchmarks using a weighted mean which places more importance on those
benchmarks with relatively larger certificates and checking times.  We
use as weight for each program its actual 
checking
time.  
We believe
that this weighted mean is more informative than the arithmetic mean,
since, for example, doubling the speed in which a large and complex
program is checked
is more relevant than achieving this for
small, simple programs.  As mentioned before, the efficiency of the
checker for reduced certificates is very similar to that of \checkerA\
(the overall slowdown is 0.99).

%% file: tablaN2.tex
\begin{table}[t]
\centering
\begin{tabular}{lrrcrrrc}\hline\hline
Program & 
Source & ByteC & BC/S & FCert &  RCert & F/R &  R/S   \\ \hline \hline
aiakl & 1555  & 3817 & 2.455 & 3090 & 1616 & 1.912 & 1.039\\ 
bid & 4945  & 10376 & 2.098 & 5939 & 883 & 6.726 & 0.179\\ 
browse & 2589  & 8492 & 3.280 & 1661 & 941 & 1.765 & 0.363\\ 
deriv & 957  & 4221 & 4.411 & 288 & 288 & 1.000 & 0.301\\ 
grammar & 1598  & 3182 & 1.991 & 1259 & 40 & 31.475 & 0.025\\ 
hanoiapp & 1172  & 2264 & 1.932 & 2325 & 880 & 2.642 & 0.751\\ 
occur & 1367  & 6919 & 5.061 & 1098 & 666 & 1.649 & 0.487\\ 
progeom & 1619  & 3570 & 2.205 & 2148 & 40 & 53.700 & 0.025\\ 
qsortapp & 664  & 1176 & 1.771 & 2355 & 650 & 3.623 & 0.979\\ 
query & 2090  & 8818 & 4.219 & 531 & 40 & 13.275 & 0.019\\ 
rdtok & 13704  & 15423 & 1.125 & 6533 & 2659 & 2.457 & 0.194\\ 
rectoy & 154  & 140 & 0.909 & 167 & 40 & 4.175 & 0.260\\
serialize & 987  & 3801 & 3.851 & 1779 & 1129 & 1.576 & 1.144\\ 
zebra & 2284  & 5396 & 2.363 & 4058 & 40 & 101.450 & 0.018\\ 
\hline \hline
 Overall & \multicolumn{2}{c}{} & 2.17  & \multicolumn{2}{c}{} & 3.35 & 0.28 \\ \hline\hline

\end{tabular}
\label{tab:tableCert}
\caption{Size of Reduced and Full Certificates}
\end{table}

%% file: tablaN1alt.tex
\begin{table}[t]
\centering
\begin{tabular}{lrrr}
\hline \hline
Program & C$_F$ & C$_R$ & C$_F$/C$_R$  \\ \hline \hline
aiakl & 85 & 86 &0.986 \\ 
bid & 46 & 48 &0.959 \\ 
browse & 20 & 20 &0.990 \\ 
deriv & 28 & 27 &1.038 \\
grammar & 14 & 14 &1.014 \\ 
hanoiapp & 31 & 30 &1.033 \\
occur & 18 & 20 &0.911 \\ 
progeom & 17 & 16 &1.012 \\ 
qsortapp & 24 & 19 &1.290 \\ 
query & 13 & 14 &0.917 \\ 
rdtok & 59 & 56 &1.061 \\ 
rectoy & 8 & 9 &0.909 \\ 
serialize & 27 & 30 &0.875 \\ 
zebra & 125 & 129 &0.969 \\ \hline
\hline
Overall& & & 0.99 \\ \hline\hline
\end{tabular}
\label{tab:tableAnCh}
\caption{Comparison of Checking Times}
\end{table}

%% file: related.tex

\section{Related Work}

A detailed comparison of the technique of ACC with related methods can
be found in \cite{ai-safety-ngc07}. In this section, we focus only
on work related to certificate size reduction in PCC.
The common idea in order to compress a certificate in the PCC scheme
is to store only the analysis information which the checker is not
able to reproduce by itself \cite{JVM03}. In the field of abstract
interpretation, this is known as \emph{fixpoint compression} and it is
being used in different contexts and tools.  For instance, in the
Astr\'ee analyzer \cite{CCFMMMR05} designed to detect runtime errors
in programs written in C, only one abstract element by head of loop is
kept for memory usage purposes. Our solution is an improvement in the
sense that some of these elements many not need to be included in the
certificate (i.e., if they are not relevant). In other words, some
loops do not require iteration to reach the fixpoint and our technique
detects this.

With our same purpose of reducing the size of certificates, Necula and
Lee~\cite{NeculaLee98a} designed a variant of the Edinburgh Logical
Framework LF \cite{HHP93}, called ${\rm LF}_i$, in which certificates
(or proofs) discard part of the information that is redundant or that
can be easily synthesized.  ${\rm LF}_i$ inherits from LF the
possibility of encoding several logics in a natural way but avoiding
the high degree of redundancy proper of the LF representation of
proofs.  In the producer side, the original certificate is an LF proof
to which a representation algorithm is applied. 
On the
consumer side, ${\rm LF}_i$ proofs are validated by using a one pass
LF type checker which is able to reconstruct on the fly the missing
parts of the proof in one pass.  Experimental results for a concrete
implementation reveal an important reduction on the size of
certificates (w.r.t. LF representation proofs) and on the checking
time.  Although this work attacks the same problem as ours the
underlying techniques used are clearly different.  Furthermore, our
certificates may be considered minimal, whereas
in~\cite{NeculaLee98a}, redundant information is still left in the
certificates in order to guarantee a more efficient behaviour of the
type checker.

A further step is taken in Oracle-based PCC~\cite{NeculaRahul01}. This
is a variation of the PCC idea that allows the size of proofs
accompanying the code to adapt to the complexity of the property being
checked such that when PCC is used to verify relatively simple
properties such as type safety, the essential information contained in
a proof is significantly smaller than the entire proof. The proof as
an oracle is implemented as a stream of bits aimed at resolving the
non-deterministic interpretation choices. Although the underlying
representations and techniques are different from ours, we share with
this work the purpose of reducing the size of certificates by
providing the checker with the minimal information it requires to
perform a proof and the genericity which allows both techniques to
deal with different kinds of properties beyond types.

The general idea of certificate size reduction has also been deployed
in lightweight bytecode verification (LBV) \cite{RR98,Ros03}. LBV is a
practical PCC approach to Java Bytecode Verification \cite{JVM03}
applied to the KVM (an embedded variant of the JVM). The idea is that
the type-based bytecode verification is split in two phases, where the
producer first computes the certificate by means of a type-based
dataflow analyzer and then the consumer simply checks that the types
provided in the code certificate are valid.  As in our case, the
second phase can be done in a single, linear pass over the bytecode.
However, LBV is  limited to types while ACC generalizes it to
arbitrary domains. Also, ACC deals with multivariance with the
associated accuracy gains (while LBV is monovariant).  Regarding the
reduction of certificate size, our work characterizes precisely the
minimal information that can be sent for a generic algorithm not tied
to any particular graph traversal strategy.  While the original notion
of certificate in \cite{RR98} includes the complete entry solution
with respect to each basic block, \cite{Ros03} reduces certificates by
sending information only for ``backward'' jumps.  As we have seen
through our running example, \cite{Ros03} sends information for all
such backward jumps while our proposal carries the reduction further
because it includes only the analysis information of those calls in
the analysis graph whose answers have been \emph{updated}, including
both branching and non-branching instructions. We believe that our
notion of reduced certificate could also be used within Rose's
framework.

As a final remark, the main ideas in ACC showed in
Equations~\ref{eq:2} and \ref{eq:4} in
Section~\ref{sec:basics-abstr-carry} have been the basis to build a
PCC architecture based on \emph{certified} abstract interpretation in
\cite{BeJePi06AIPCC}.  Therefore, this proposal is built on the basics
of ACC for certificate generation and checking, but
relies on a \emph{certified} checker specified in Coq~\cite{coq} in
order to reduce the trusted computing base. In contrast to our
framework, this work is restricted to safety properties which hold for
all states and, for now, it has only been implemented for a particular
abstract domain.